\documentclass[%
 reprint,
 amsmath,amssymb,
 aps,
 prl,
 twocolumn,
 bibnotes,
 floatfix
]{revtex4-2}
\usepackage{graphicx}
\usepackage{dcolumn}
\usepackage{bm}
\usepackage{float}
\usepackage{supertabular}
\usepackage{makecell}
\usepackage{booktabs}
\usepackage{comment}
\usepackage{color}
\usepackage{import}
\usepackage{multirow}
\usepackage{textcomp}
\usepackage{setspace}
\usepackage{longtable}
\usepackage[linesnumbered, ruled]{algorithm2e}
\usepackage[colorlinks,citecolor=blue,linkcolor=blue]{hyperref}

\begin{document}
\title{Experimental Asynchronous Measurement-Device-Independent Quantum Cryptographic Conferencing}
\author{$\text{Yifeng Du}^1$}
\author{$\text{Yang Hu}^1$}
\author{$\text{Yufeng Liu}^1$}
\author{$\text{Wenhan Yan}^1$}
\author{$\text{Jinghao Zhang}^1$}
\author{$\text{Shining Zhu}^1$}
\author{$\text{Xiao-Song Ma}^{1,2,3,4,5,6}$}
\email{xiaosong.ma@nju.edu.cn}
\affiliation{$^1$National Laboratory of Solid State Microstructures, School of Physics,Collaborative Innovation Center of Advanced Microstructures, Nanjing University, Nanjing 210093, China\\
$^2$Jiangsu Physical Science Research Center, Nanjing 210093, China\\
$^3$Shishan Laboratory, Nanjing University, Suzhou 215163, China\\
$^4$Jiangsu Key Laboratory of Quantum Information Science and Technology, Nanjing 210093, China\\
$^5$Synergetic Innovation Center of Quantum Information and Quantum Physics, University of Science and Technology of China, Hefei, Anhui 230026, China\\
$^6$Hefei National Laboratory, Hefei 230088, China
}

\begin{abstract}

The quantum cryptographic conferencing (QCC) protocol, which distributes identical secure keys to user groups, is a crucial component of the quantum network. Previous experimental works have implemented the measurement-device-independent (MDI) QCC, of which the key rate in an N-user network scales as $R\sim O(\eta^N)$. Building on the MDI QCC protocol, the asynchronous MDI QCC protocol theoretically integrates the mode-pairing scheme into QCC, significantly boosting the key rate to $R\sim O(\eta)$ in the ideal case, which is independent of the number of users, thus demonstrating greater application potential. Experimentally, in this Letter, we implement the three-user asynchronous MDI QCC network without phase locking by adopting the fast Fourier transform-based frequency difference estimation and the phase drift compensation technique. Finally, we achieve a key rate of about $4.470\times 10^{-9}$ bits per pulse under a maximum overall loss of about 59.6 dB. This Letter provides a scalable solution for the development of large-scale quantum communication networks in the future.
\end{abstract}

\maketitle

\section{Introduction}
Quantum networks aim to connect users through quantum communication protocols\cite{qnet_review_1,qnet_review_2}. Founded on the fundamental principles of quantum mechanics, quantum networks offer advantages that are unattainable with classical communication. They can operate in parallel with the traditional Internet, thereby enhancing communication performance. Within this promising blueprint, the quantum key distribution (QKD) protocol \cite{BB84,qkd_review_4,qkd_review_1,qkd_review_2,qkd_review_3,qkd_review_5} endowed with the property of information-theoretic unconditional security serves as a critical component. Traditional point-to-point QKD protocols between two users have achieved substantial progress and development both theoretically \cite{decoy1,decoy2,decoy3,diqkd,Pirandola2012,mdi_qkd,tf_qkd,mp_qkd_2,mp_qkd_1} and experimentally \cite{1120km_qkd,chen_integrated_2021,110M_qkd,1000km_qkd}, moving toward real-world practical application. Meanwhile, the research on multiuser QKD networks develops rapidly \cite{qkd_network_2,qkd_network_3,qkd_network_1,yan_measurement-device-independent_2025}. Building on this foundation, the quantum cryptographic conferencing (QCC) protocol, also known as the quantum conference key agreement protocol \cite{qcc_concept_1,qcc_concept_2,qcc_review}, which extends two-user secure communication to multiuser scenarios, is key to further expanding the scale of quantum communication networks and improving their functionality.

QCC protocols distribute secure keys among communicating users via genuine multipartite entanglement states \cite{qcc_gme_2,qcc_gme_3,qcc_gme_1}. To date, the QCC protocols proposed theoretically can be categorized into entanglement-based (EB) protocols \cite{qcc_eb_theory_2,qcc_eb_theory_1} and time-reversal protocols \cite{pol_qcc,w_qcc_1,pm_qcc,cow_qcc_2,cow_qcc_1,cow_qcc_3,w_qcc_2,lu2024repeaterlike,xie2024multifield}. In EB protocols, a source generates multiphoton entanglement states and transmits them to users for measurement. Previous relevant experimental works have realized the multiuser QCC communication networks by preparing by preparing four-photon Greenberger-Horne-Zeilinger (GHZ) states \cite{qcc_eb_exp_1}, and six-photon graph states with more complex entanglement structures \cite{qcc_eb_exp_2}. However, the complexity and low generation rate of multiphoton entanglement sources limit the practicality of EB protocols. 

Time-reversal protocols, by contrast, establish quantum correlations among users through projective measurement of genuine multipartite entanglement states. Some of these protocols possess the measurement-device-independent (MDI) property, which makes them immune to all attacks targeting the detection node and, thus, ensures higher security. Furthermore, the user node and detection node of MDI QCC systems are compatible with mature MDI QKD systems, making the MDI QCC protocol easier to implement in practical applications. Theoretically, various MDI QCC protocols have been successively proposed, including polarization-encoding protocol \cite{pol_qcc}, phase-matching protocol \cite{pm_qcc}, asynchronous MDI (AMDI) protocol \cite{lu2024repeaterlike}, multi-field protocol \cite{xie2024multifield}, and W-state-based protocols \cite{w_qcc_1,w_qcc_2}. In particular, the AMDI QCC protocol extends the mode-pairing (MP) scheme from QKD to QCC, improving the original MDI QCC protocol, which requires multiphoton coincidence counting, into a single-photon protocol capable of breaking the Pirandola-Laurenza-Ottaviani-Banchi (PLOB) bound \cite{pirandola_fundamental_2017} with a key rate of $R\sim O(\eta)$ in the ideal case. Meanwhile, the AMDI QCC also eliminates the need for complex phase tracking technology. In terms of experimental research, recent works have successfully implemented the three-user polarization-encoding MDI QCC network \cite{PhysRevLett.133.210803,du_experimental_2025}, and phase-matching QCC network \cite{pm-qcc-exp}, verifying the feasibility of the these protocols. Nevertheless, the low multiphoton coincidence counting rate under high channel loss restricts the large-scale application of these protocols.

Building on previous works of MP QKD \cite{mp_zhu_experimental_2023,mp_zhou_experimental_2023,mp_zhu_field_2024,mp_zhang_experimental_2025,mp_shao_high-rate_2025,mp_lu_experimental_2025} and MDI QCC \cite{PhysRevLett.133.210803,du_experimental_2025}, in this Letter, we construct a three-user communication network and successfully implement the AMDI QCC protocol \cite{lu2024repeaterlike} without phase locking. Through the fast Fourier transform (FFT)-based frequency difference estimation \cite{li_twin-field_2023,mp_zhang_experimental_2025,mp_shao_high-rate_2025,mp_lu_experimental_2025} and the phase drift postcompensation technique for the multipath interferometer, we achieve a secure key rate of about $3.940\times10^{-8}$, $3.937\times10^{-8}$, and $4.470\times10^{-9}$ bits per pulse with an overall system loss of three users of about 39.3, 48.6, and 59.6 dB, respectively. Compared to the MDI QCC experiments \cite{PhysRevLett.133.210803,du_experimental_2025}, our work significantly increases the overall loss tolerance of the communication system by more than 30 dB and enhances the communication distance and secure key rate of the MDI QCC protocol through a simple approach. Our Letter not only experimentally validates the feasibility of asynchronous mode-pairing schemes in multiuser quantum communication, but also paves the way for the practical implementation of future large-scale quantum networks.

\section{Protocol}
In the $N$-user AMDI QCC protocol, each user $U_i$ independently prepares a phase-randomized coherent state $\left|\sqrt{k_i}\text{e}^{\text{i}\varphi_i} \right\rangle$, where the coherent state intensity $k_i$ is randomly selected from the signal state $\mu$, decoy state $\nu$ and vacuum state $o$; $\varphi_i$ denotes a random phase. The optical pulses are sent by users to the detection node, where each pulse interferes with the pulse from the adjacent user at the beam splitters. If there is exactly one detector response within the time window, the detection node records and publicly announces information, including the single-count event and the responding detector. Following multiple repetitions of the above process, users pair the single-count events from $N$ detectors of different branches. Since the AMDI-QCC protocol generates keys through single-photon counting events, its total number of pairs scales approximately linearly with the single-user channel transmittance $\eta$, leading to an asymptotic key rate scaling as $O(\eta)$ in the ideal case. Based on the corresponding total intensity $k_i^{\text{tot}}=k^e_i+k^l_i$ and encoded random phase difference $\theta_i=\varphi_i^l-\varphi_i^e$ of the paired early and late pulses announced by users, they determine the encoding basis and bit value, and calculate the corresponding quantum bit error rate (QBER). Specifically, the pairs with $k_i^{\text{tot}}=\mu$ will be assigned to $Z$-basis events for key distillation, and the events with $k_i^{\text{tot}}=2\nu$ and the total encoding phase difference of all users $\sum_{i=1}^N{\theta_i} \mod 2\pi=0$ or $\pi$ will be assigned to $X$-basis. A secure conference key can be obtained after estimating parameters via the decoy-state method and performing classical postprocessing \cite{lu2024repeaterlike}.

For a communication network with $N=3$ users, the final secure key length under finite size is given by \cite{lu2024repeaterlike}:
\begin{eqnarray}\label{eq1}
        L_{\text{min}}&=&\underline{Y}^Z_{111}\left[1-H_2\left(\overline{e}^{PZ}_{111} \right) \right]-\text{max}\left[H_2\left(e^Z_{AB,AC,BC}\right)\right]fn_{Z}\nonumber\\
        &\;&-\log_2\left(\frac{4}{\varepsilon_{\text{cor}}}\right)-2\log_2\left(\frac{2}{\varepsilon'\hat{\varepsilon}}\right)-2\log_2\left(\frac{1}{2\varepsilon_{\text{PA}}}\right),
\end{eqnarray}
where $\underline{Y}^Z_{111}$ and $\overline{e}^{PZ}_{111}$ are the lower bound of the number of $Z$-basis single-photon pairs and the upper bound of the corresponding phase error rate, respectively; $e^Z_{AB, AC, BC}$ are the $Z$-basis QBERs between users $AB$, $AC$, and $BC$, respectively\textemdash i.e., the probabilities that the bit values of the Z-basis pairs of the two users are inconsistent; $\text{max}\left[H_2\left(e^Z_{AB,AC,BC}\right)\right]=\max\left\{H_2\left(e^Z_{AB}\right),H_2\left(e^Z_{BC}\right),H_2\left(e^Z_{AC}\right)\right\}$ and $H_2(x)=-x\log_2x-(1-x)\log_2(1-x)$ is the binary Shannon entropy function; $n_{Z}$ is total number of $Z$-basis pairs; $f$ is the error correction efficiency; $\varepsilon_{\text{cor}}$, $\varepsilon'$, $\hat{\varepsilon}$ and $\varepsilon_{\text{PA}}$ are security parameters.

\section{Experimental Setup}

\begin{figure*}
\centering
\includegraphics[width=\linewidth]{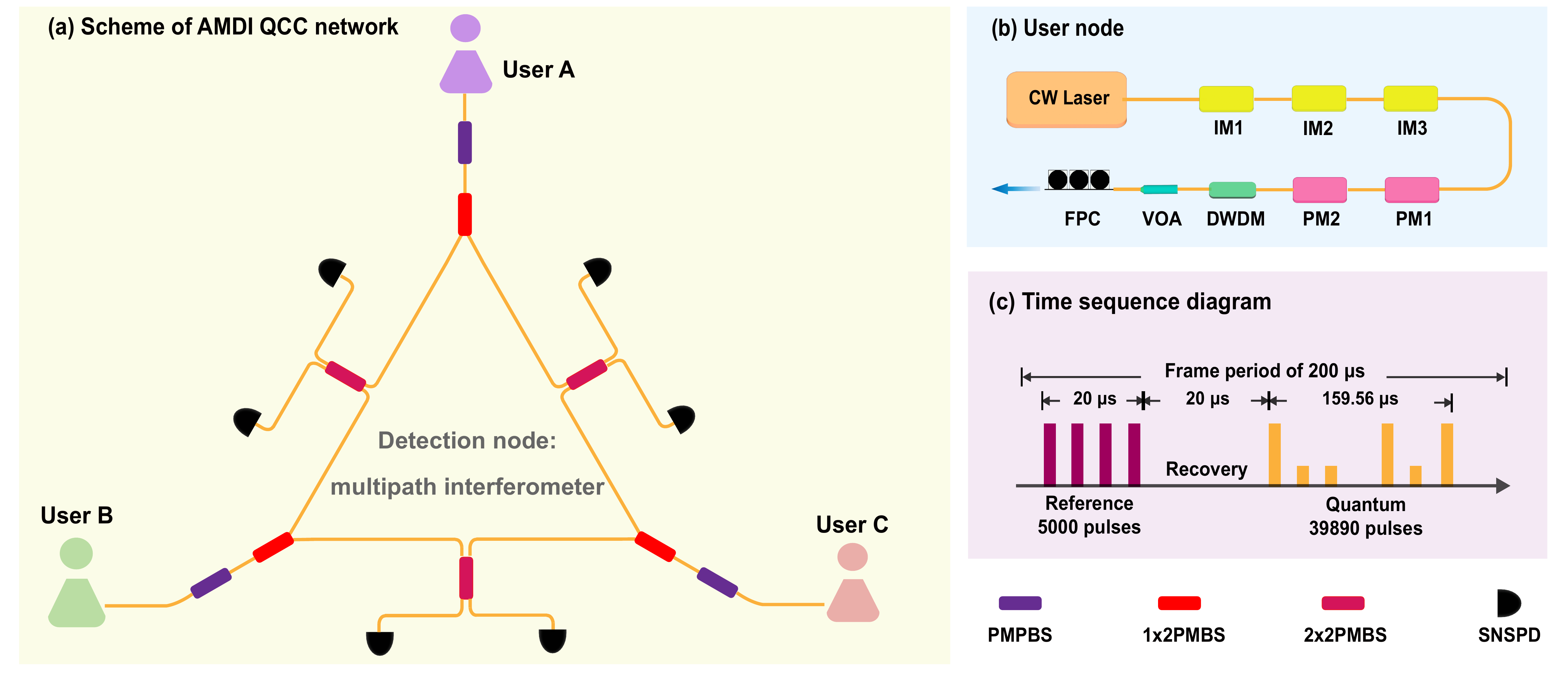}
\caption{\label{fig:scheme}Experimental setup of AMDI QCC. (a) Scheme of the AMDI QCC network. The user transmits the encoded pulses to the multipath interferometer at the detection node for measurement. (b) The user node. Each user employs an independent free-running laser as the light source. The encoder section consists of three cascaded intensity modulators and two phase modulators, which perform decoy state and 16-slice random phase modulation, respectively. (c) The time sequence diagram of pulses prepared by users. The complete pulse sequence comprises the reference pulses for frequency difference and interferometer phase drift estimation, the quantum pulses for key generation, and the recovery region between reference and quantum pulses. Abbreviations of the components: BS, beam splitter; IM, intensity modulator; PM, phase modulator; DWDM, dense wavelength division multiplexer; VOA, variable optical attenuator; FPC, fiber polarization controller; PMPBS, polarization-maintaining polarization beam splitter; PMBS, polarization-maintaining beam splitter; SNSPD, superconducting nanowire single-photon detector.}
\end{figure*}

The experimental setup of the three-user communication network constructed in this work is illustrated in FIG. \ref{fig:scheme}. In the user nodes, three independent free-running continuous-wave lasers with a central wavelength of about 1536.61 nm and a nominal linewidth of less than about 1 kHz are used as the light source. In the encoder section, each user first employs three cascaded electro-optic intensity modulators (IMs) to modulate the continuous light into an optical pulse series with a repetition frequency of 250 MHz and a pulse width of about 260 ps in the key generation experiments. Each complete pulse sequence contains 5000 reference light pulses with approximately the same intensity as the signal state and 39 890 quantum light pulses for key generation. Besides, there is a recovery region of about 20 $\mu$s between the reference pulses and the quantum pulses. In the phase-encoding process following optical pulse generation, according to the previous works \cite{Cao_2015,PhysRevResearch.2.043304,PhysRevApplied.16.034017,1000km_qkd,mp_zhu_experimental_2023,mp_zhou_experimental_2023}, each user implements discrete random phase encoding with the slice number of $M=16$ using two electro-optic phase modulators (PMs), i.e., $\varphi_i=2\pi M_i/M$, where $M_i\in\{0,1,\dots, 15\}$. Specifically, users apply electrical signals with amplitudes $\{0, V_{\pi}\}$ to the first phase modulator, and electrical signals of the same pulse width with amplitudes $\{0, V_{\pi/8}, \dots, V_{7\pi/8}\}$ to the second phase modulator, in which $V_{\pi}$ is the half-wave voltage of the PMs. The pulse width of the electrical signals for phase modulation is about 1 ns.

After completing the encoding process, users filter the optical pulses with dense wavelength-division multiplexers (DWDMs), attenuate them to the single-photon level, and simulate fiber link losses using variable attenuators (VOAs) when transmitting the pulses to the detection node. Then the fiber polarization controllers (FPCs) and the fiber polarization maintaining polarization beam splitters (PMPBSs) work in coordination to align their polarization states. In the interferometer, the input light is first split into two parts by a $1\times2$ polarization-maintaining beam splitter (PMBS), and each part is interfered with photons from the other two users at a $2\times2$ PMBS, respectively. The split ratio of all beam splitters in the interferometer is about 50:50. The average insertion loss of the multipath interferometer is about 3.29 dB. Superconducting nanowire single-photon detectors (SNSPDs) are employed to detect the output photons of the interferometer, exhibiting the average dark count rate of about 42 Hz, and the average detection efficiency of about 75.9\%. Finally, a time-to-digital converter is used to record the time tag and channel of the detection events.

\begin{figure}[h]
\includegraphics[width=\linewidth]{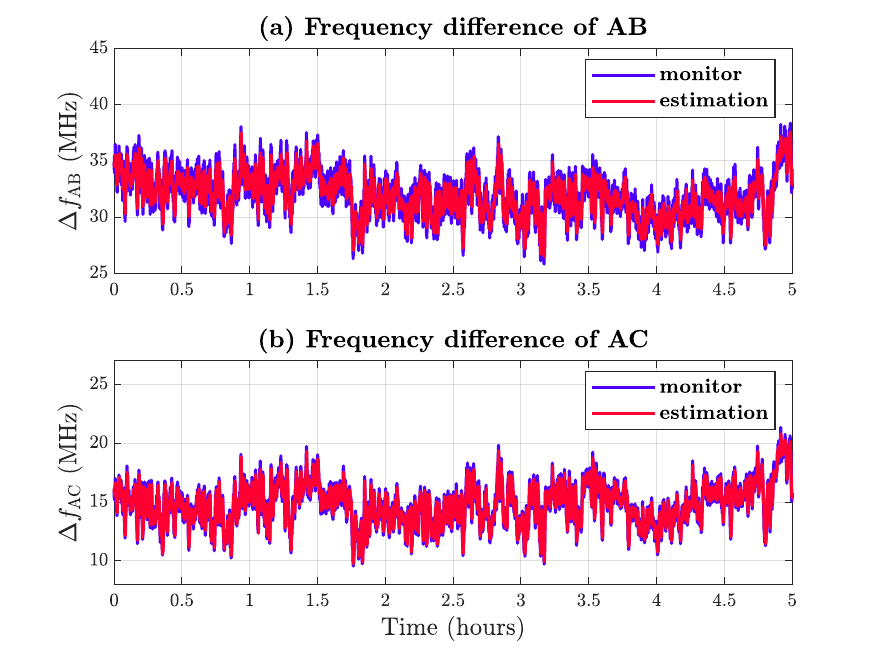}
\caption{\label{fig:freq}Frequency differences estimation test results and beating frequency measurement results between (a) users $AB$, and (b) users $AC$ in 5 h. 
}
\end{figure}

\begin{figure*}
\includegraphics[width=1.0\linewidth]{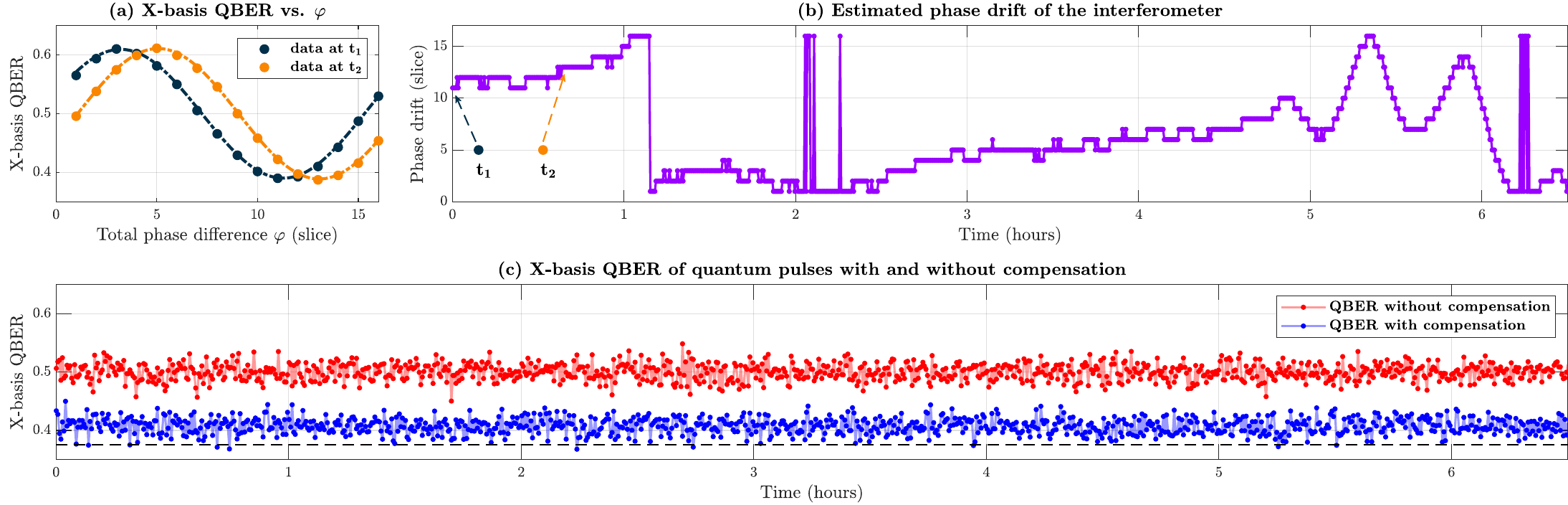}
\caption{\label{fig:phase_drift}Phase drift compensation and $X$-basis QBER. (a) Relationship between the $X$-basis QBER of the reference pulses and the total phase difference of the paired pulses after frequency difference compensation. The visibility of GHZ-HOM interference obtained by fitting the experimental data is about 0.22. The horizontal shift between the dark blue and orange cosine curves measured at $t_1$ and $t_2$ with a time interval of about 0.65 h indicates the drift of the additional phase $\Delta\varphi$. (b) Estimated phase drift of the interferometer. We regard the phase slice value corresponding to the minimum QBER point in (a) as the phase shift estimation value. In the calculations, we compute the phase values within the range of 0 to $2\pi$, and, thus, an abrupt change occurs in the phase shift values when the phase exceeds this range. (c) $X$-basis QBER of quantum pulses with and without phase compensation. After compensating for the frequency difference and phase drift in the test, the $X$-basis QBER is reduced from about 50.02\% to about 40.76\%, with a theoretical minimum value of about 37.50\% (black dashed line).}
\end{figure*}

\section{Estimation of frequency differences}
In AMDI QCC, the sifting and key mapping of $X$-basis pairs depend on the encoding phase difference between paired early and late pulses. However, in the experiment with independent free-running lasers, the frequency differences between the light sources introduce an additional phase difference that needs to be estimated and compensated for the paired pulses to decrease $X$-basis QBER. In the experiment, we pair reference pulses pairwise according to the method described in the MP QKD protocol and analyze the spectrum of the X-basis QBER using the FFT algorithm to estimate the frequency differences between the two users \cite{li_twin-field_2023,mp_zhang_experimental_2025,mp_shao_high-rate_2025,mp_lu_experimental_2025}. As shown in FIG. \ref{fig:freq}, in the test, we estimate the frequency difference about every 0.4 seconds and monitor the beating frequency between users $AB$ and $AC$ via an oscilloscope over a period of about 5 h for comparison, which is similar to the method given in \cite{mp_shao_high-rate_2025}. The results indicate that the estimation values are in good agreement with the measured values, demonstrating that our method can accurately estimate frequency differences. Based on the estimated frequency difference and the time interval between paired single counts, we can calculate the corresponding phase shift compensation value $\theta_{\Delta f}$ by the method given in the Supplemental Material in detail \footnote{\label{footnote}See Supplemental Material for the detailed analysis of the frequency difference estimation method (Sec. S3)}. 

\section{Phase drift post-compensation}

On the other hand, the inherent phase drift of the multipath interferometer also has a significant impact on the $X$-basis QBER. Ideally, supposing that the pairing time interval is sufficiently short such that the phase drift induced by the long-distance fiber links can be neglected, then the total phase difference of the paired pulses is $\varphi=\sum_{i=1}^N{\theta_i}+\theta_{\Delta f}$ and the $X$-basis QBER will be $E_X=(1-V\cos\varphi)/2$, in which $\sum_{i=1}^N{\theta_i}$ is the total encoding phase difference, $\theta_{\Delta f}$ is the phase shift introduced by frequency difference, and $V$ is the visibility of three-photon GHZ-HOM (Hong-Ou-Mandel) interference \cite{du_experimental_2025}. The dependence of the $X$-basis QBER $E_X$ on the pairing time interval is included in the $\theta_{\Delta f}$ term.
As the filter criterion for the $X$-basis pairs, when $\varphi=0$ or $\pi$, the QBER reaches its extremum. However, in practice, there is a relative phase between two subpulses generated by the $1\times2$ BSs in the interferometer, which introduces an additional phase shift $\Delta\varphi$, resulting in the $X$-basis QBER becoming $E_X=[1-V\cos(\varphi-\Delta\varphi)]/2$ as shown in FIG. \ref{fig:phase_drift}(a). The additional phase shift $\Delta\varphi$ in the all-fiber interferometer is prone to rapid drifting due to environmental disturbances, which causes the long-term average $X$-basis QBER to approach 50\%.
To solve this problem, in the experiment, we place the interferometer in an insulated box to maintain its phase relatively stable for over 20 s. Then we estimate and compensate for the phase drift of the interferometer during postprocessing. We perform three-photon pairing on the reference pulses after compensating for the frequency difference and regard the phase slice value $\theta_{\mathrm{min}}$ corresponding to the minimum QBER as the phase drift estimation value. During the key mapping process in the protocol, considering the phase drift of the interferometer, we adjust the filter criteria determining the X-basis pair to $\left(\sum_{i=1}^N{\theta_i}+\theta_{\Delta f}\right) \mod 2\pi\in\{\theta_{\text{min}}, \theta_{\text{min}}+\pi\}$. In the experiment, we perform the above process on about every 20 s of data. As shown in FIG. \ref{fig:phase_drift}(c), by combining frequency difference estimation and phase drift postcompensation of the interferometer, we reduce the $X$-basis QBER of the reference pulses from about 49.80\% to about 39.14\%, while simultaneously reducing the $X$-basis QBER of the quantum pulses from about 50.02\% to about 40.76\%.

\section{Key generation results}
\begin{figure}[htbp]
	\centering
    \includegraphics[width=1.0\linewidth]{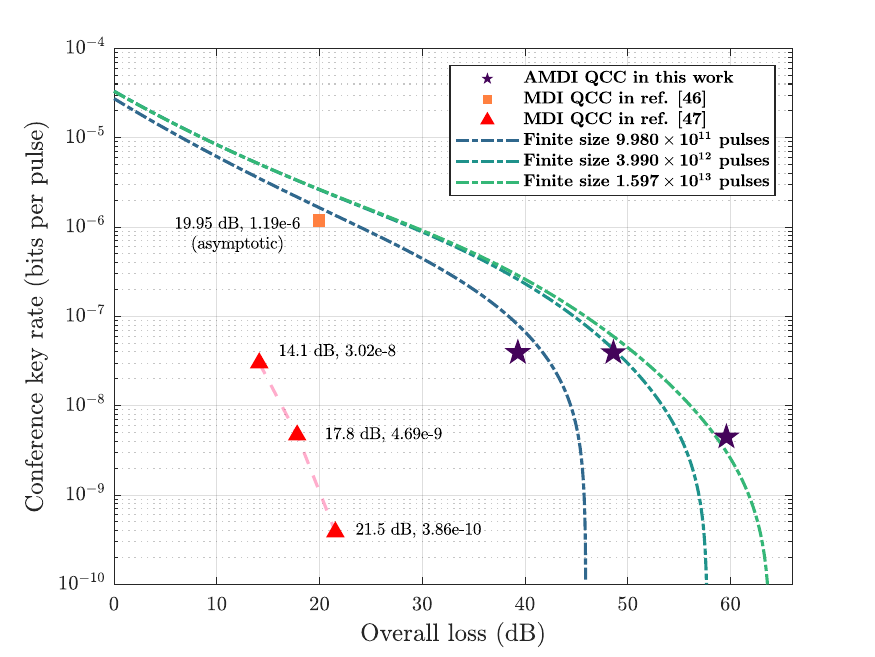}
	\caption{\label{fig:rate}Key generation results of the AMDI QCC in this Letter, and MDI QCC in \cite{PhysRevLett.133.210803,du_experimental_2025}. The dashed curves in the figure represent theoretical key rates corresponding to scenarios with data accumulation time of about $5\times 10^3$, $2\times 10^4$, and $8\times 10^4$ s, corresponding to the total number of transmitted quantum pulses of about $9.980\times 10^{11}$, $3.990\times 10^{12}$, and $1.597\times 10^{13}$, respectively. The parameters used in the simulation are given in Supplemental Material [\hyperref[footnote]{59}]. For comparison, the experimental results of the prior works cited in the figure are calculated based on the data provided in those references.}
\end{figure}

We perform key generation experiments under conditions with the all-user overall loss including insertion loss and detection efficiency of about 39.3, 48.6, and 59.6 dB. In the above cases, we accumulate about $5\times10^3$, $2\times10^4$, and $8\times 10^4$ s of data, respectively. The encoding parameters used in the experiment are set as $\mu=0.286$, $\nu=0.026$, $p_{\mu}=0.1008$, $p_{\nu}=0.5398$, and the maximum pairing time interval $t_{\text{max}}=3$ $\mu$s. 
The final secure key rates are about $3.940\times10^{-8}$, $3.937\times10^{-8}$, and $4.470\times 10^{-9}$ bpp with an all-user overall loss of about 39.3, 48.6, and 59.6 dB, respectively. The key generation results in this Letter and other relevant works are shown in FIG. \ref{fig:rate}. 
Compared with the previous MDI QCC experimental work \cite{PhysRevLett.133.210803,du_experimental_2025}, the system’s tolerable channel loss and secure key rate have been significantly improved from 21.5 to about 59.6 dB without a noticeable increase in experimental complexity. Although our experiment does not achieve the ideal $O(\eta)$ key rate and break the PLOB bound, the experimental results at $t_{\mathrm{max}}=\infty$, i.e., with no restriction on the pairing time interval, verify the linear relationship between the total number of pairs and the channel transmittance $\eta$, as detailed in Sec. 4.4 of Supplemental Material [{\hypersetup{linkcolor=blue}\hyperref[footnote]{59}].

\section{Discussion}
In this Letter, we have successfully implemented the AMDI QCC protocol without phase locking. Ideally, the key rate of the AMDI QCC protocol in the $N$-user network scales as $R\sim O(\eta)$, which is independent of the number of users and, thus, provides a more scalable scheme for the multiuser quantum network. Benefiting from the theoretical advancements of the protocol, the communication distance of the system in this work has been greatly improved compared with the previous experimental works on the MDI QCC protocol \cite{PhysRevLett.133.210803,du_experimental_2025}. With the rapid development of related technologies such as on-chip integrated multiwavelength light sources \cite{kippenberg_dissipative_2018,kues_quantum_2019,shen_integrated_2020,moille_kerr-induced_2023,huang_massively_2025,yan_ten-channel_2025}, modulators \cite{wang_integrated_2018} and detectors \cite{grunenfelder_fast_2023}, it is expected that, in the future, we can realize intercity quantum multiparty communication with a distance from single-user node to detection node exceeding 100 km over optical fiber links, and lay a solid foundation for the realization of large-scale and scalable quantum networks. 

\section{Acknowledgment}
This research was supported by HFNL Self-Deployed Project (No. ZB2026020200), the Natural Science Foundation of Jiangsu Province (Grants No. BK20240006 and No. BK20233001), Quantum Science and Technology-National Science and Technology Major Project (Grants No. 2021ZD0300700 and No. 2021ZD0301500), Fundamental and Interdisciplinary Disciplines BreakthroughPlan of theMinistry of Education of China (No. JYB2025XDXM112), and Nanjing University-China Mobile Communications Group Co., Ltd. Joint Institute.
\section{Data availability}
The data that support the findings of this article are openly available \footnote{\url{https://github.com/NJU-Malab/AMDI-QCC}}.

\newpage    
\onecolumngrid
\section*{Supplemental Materials}
\setcounter{section}{2} 
\setcounter{figure}{0} 
\setcounter{equation}{0}
\renewcommand*{\thefigure}{S\arabic{figure}}
\renewcommand*{\thetable}{S\arabic{table}}
\renewcommand*{\thesection}{S\arabic{section}}
\renewcommand*{\theequation}{S\arabic{equation}}
\counterwithout{equation}{section}
\numberwithin{equation}{section}

\section{S1 Pairing Algorithm of AMDI QCC in the experiment}
\begin{algorithm}[H]
\setstretch{0.87}
\caption{Pairing Algorithm of AMDI QCC in the experiment} 
    \label{alg:AOS}
    \SetKw{length}{length}
    \SetKw{and}{and}
    \SetKw{isNoEmptyEvents}{isNoEmptyEvents}
    \SetKw{or}{or}
    \SetKw{break}{break}

        \KwIn{The set $S=\{E_1, E_2, \cdots, E_N\}$ of effective single-click events, where the $i$th event $E_i=\{C_i, T_i\}$ has properties $C_i$ (the detector channel) and $T_i$ (the click time); maximum pairing time $t_{\text{max}}$} 
        
        \KwOut{The set $P=\{P_1, P_2, \cdots\}$ of pairing events, where the $j$th pair $P_j=\{p_1^j, p_2^j, \cdots, p_n^j\}$ consists of single-click events from $n$ different branches.}    
        
        Initialization: loop index $i=1$; iterate starting index $k_{\text{start}}=1$; temporarily unpaired events index $k_{\text{pairing}}=N$ and $k_{\text{unpairing}}=N$; pair number index $k_p=1$; temporary empty register set $A$; the number of events in the register $n_{\text{temp}}$\;

        \While{$i<N-1$}
        {
            \For{$j=k_{\mathrm{start}}$ \KwTo $N$}
            {
                identify click branch $B_j$ from click detector channel $E_j.C_j$\;
                \eIf{$j=k_{\mathrm{start}}$}
                {
                    append the event $E_j$ into $A$: $A[B_j]=E_j$\;
                    set starting timetag: $t_{\mathrm{start}}=E_j.T_j$
                }
                {
                    \eIf{$A[B_j]=\varnothing$}
                    {
                        \eIf{$E_j.T_j-t_{\mathrm{start}}\le \Delta t_{\mathrm{max}}$ \and $j<N$}
                        {
                            append the event $E_j$ into $A$: $A[B_j]=E_j$\;
                            set temporarily unpaired events index: $k_{\text{pairing}}=\min{[k_{\text{pairing}},j]}$\;
                            count the number of events in the register: $n_{\mathrm{temp}}=n_{\mathrm{temp}}+1$\;
                            \If{$\varnothing \not\subset A[k_p]$}
                            {
                                append the single counts pair $A$ into set $P[k_p]$\;
                                update the corresponding pair index: $k_{p}=k_p+1$\;
                                set new iterate starting index $k_{\mathrm{start}}=\min{[j+1,k_{\mathrm{unpairing}}]}$\;
                                set new loop index $i=k_{\mathrm{start}}$\;
                                re-initialize the temporarily unpaired events index $k_{\mathrm{pairing}}=N$, $k_{\mathrm{unpairing}}=N$, temporary empty register set $A$, and the number of events in the register $n_{\mathrm{temp}}$\;
                                \break\;
                            }
                        }
                        {
                            \eIf{$n_{\mathrm{temp}}=1$}
                            {
                                set new iterate starting index $k_{\mathrm{start}}=k_{\mathrm{start}}+1$\;
                            }
                            {
                                set new iterate starting index $k_{\mathrm{start}}=\min{[k_{\mathrm{pairing}},k_{\mathrm{unpairing}}]}$\;
                            }
                            set new loop index $i=k_{\mathrm{start}}$\;
                            re-initialize the temporarily unpaired events index $k_{\mathrm{pairing}}=N$, $k_{\mathrm{unpairing}}=N$, temporary empty register set $A$, and the number of events in the register $n_{\mathrm{temp}}$\;
                            \break\;
                        }
                    }
                    {
                        \eIf{$E_j.T_j-t_{\mathrm{start}}\le \Delta t_{\mathrm{max}}$ \and $j<N$}
                        {
                            set temporarily unpaired events index: $k_{\mathrm{unpairing}}=\min{[k_{\mathrm{unpairing}},j]}$
                        }
                        {
                            \eIf{$n_{\mathrm{temp}}=1$}
                            {
                                set new iterate starting index $k_{\mathrm{start}}=k_{\mathrm{start}}+1$\;
                            }
                            {
                                set new iterate starting index $k_{\mathrm{start}}=\min{[k_{\mathrm{pairing}},k_{\mathrm{unpairing}}]}$\;
                            }
                            set new loop index $i=k_{\mathrm{start}}$\;
                            re-initialize the temporarily unpaired events index $k_{\mathrm{pairing}}=N$, $k_{\mathrm{unpairing}}=N$, temporary empty register set $A$, and the number of events in the register $n_{\mathrm{temp}}$\;
                            \break\;
                        }
                    }
                }
            }
        }
\end{algorithm}

In the experiment, we adopt the nearest-neighbor pairing algorithm based on \cite{lu2024repeaterlike}. When the program starts running, it takes the first single-count event as the starting point, and then searches forward for the single-count events belonging to other branches with the nearest time intervals. If the time interval between the latest single-count event and the starting point is within the $t_{\text{max}}$, then these events from $N$ different branches will be regarded as a pair, and the unpaired single-count event closest to the starting event will be set as the new starting point. Otherwise, the first unpaired event will be set as the new starting point and the above process will be repeated when the pairing is not successfully achieved within the specified time interval $t_{\text{max}}$. 

\section{S2 Key mapping and decoy state method}
\subsection{S2.1 Key mapping}
Similar to the time-bin MDI QKD protocol and the MP QKD protocol, in the AMDI QCC protocol, the pairs of early and late pulses that carry users’ encoded information are determined based on the rules illustrated in FIG. \ref{fig:s1} below after valid pairing events are obtained:

\begin{figure}[H]
	\centering
	\includegraphics[width=0.8\linewidth]{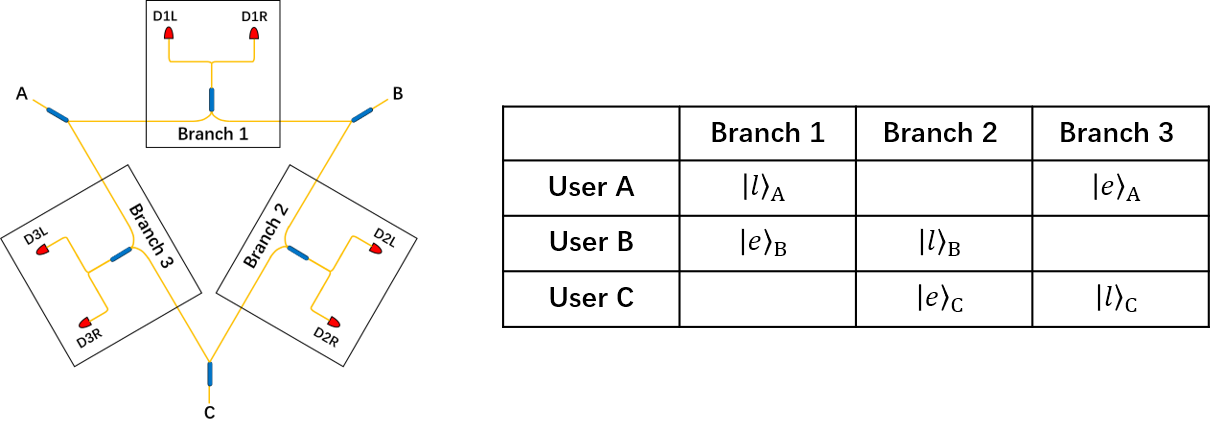}
	\caption{\label{fig:s1}Three-user AMDI QCC pulse pairing scheme. The table on the right shows the correspondence between the paired pulses of different users and the single-count events from different interferometer branches in the experiment, in which the $e$ and $l$ represent the early pulse and late pulse, respectively.}
\end{figure}

Let the intensities and phases of the early and late pulses for the three users be denoted as $k^{e(l)}_{1,2,3}$ and $\varphi^{e(l)}_{1,2,3}$, respectively. The total intensities of the early and late pulses and their phase differences are defined as $k^{\text{tot}}_{1,2,3}=k^{e}_{1,2,3}+k^{l}_{1,2,3}$, $\Delta\varphi_{1,2,3}=\varphi^{l}_{1,2,3}-\varphi^{e}_{1,2,3}$, respectively. A pairing event where the total intensities of the three users are $k_1^{\text{tot}}$, $k_2^{\text{tot}}$, and $k_3^{\text{tot}}$is represented as ${[k_1^{\text{tot}}, k_2^{\text{tot}}, k_3^{\text{tot}}]}$. The methods for determining the Z-basis, X-basis, and their corresponding bit values are as follows:

\begin{itemize}
    \item Z-basis code: In the experiment, we adopted the {with-filter} scheme proposed in \cite{lu2024repeaterlike}, which means only the pairing $([\mu,\mu,\mu])$—where the total pulse intensity corresponds to the signal state—is selected as a Z-basis event. If $k^e=\mu$ and $k^l=o$, the Z-basis bit value is 0; conversely, if $k^e=o$ and $k^l=\mu$, the Z-basis bit value is 1. A bit error occurs when the Z-basis bit values of all users are not identical. In the {without-filter} scheme, pairings such as $[\mu,\mu,\mu]$, $[\mu,\mu,\nu]$, $[\mu,\nu,\mu]$, $[\mu,\nu,\nu]$, $[\nu,\mu,\mu]$, $[\nu,\mu,\nu]$, $[\nu,\nu,\mu]$, and $[\nu,\nu,\nu]$ can all be regarded as Z-basis events, which is not adopted in our experiments.

    \item X-basis code: In the ideal case, pairs of $[2\nu, 2\nu, 2\nu]$ that satisfy $\theta^X_{\text{parity}}=(\sum_{i=1}^N{\theta_i})/\pi \mod 2\in \{0,1\}$ are selected as X-basis events. Let $r_{1,2,3}\in\{0, 1\}$ denote the response of detector $D_{L}$ or $D_{R}$ in the three output branches of the interferometer, respectively; then, pairs that satisfy $r_1\oplus r_2\oplus r_3\oplus \theta^X_{\text{parity}}=1$ are regarded as X-basis bit error events. In the key generation experiment, for the quantum pulses, the above sifting condition can be modified as $(\sum_{i=1}^N{\theta_i}+\theta_{\Delta f})\mod 2\pi\in \{\theta_{\text{min}},\theta_{\text{min}}+\pi\}$, where the $\theta_{\Delta f}$ and $\theta_{\text{min}}$ denote the phase shift introduced by frequency difference and phase drift in the interferometer as mentioned in the text. Besides, for the reference pulses, we keep all $[2\nu, 2\nu, 2\nu]$ pairs and set the pairs with $r_1\oplus r_2\oplus r_3=1$ as error events to obtain the relationship between X-basis QBER and total phase difference, such that we can estimate the phase drift $\theta_{\text{min}}$. 
\end{itemize}

\subsection{S2.2 Decoy state method}
According to the secure key length formula: 
\begin{equation}\label{eq:key_rate}
        L_{\text{min}}=\underline{Y}^Z_{111}\left[1-H_2\left(\overline{e}^{PZ}_{111} \right) \right]-\text{max}\left[H_2\left(e^Z_{AB,AC,BC}\right)\right]fn_{Z}-\log_2\left(\frac{4}{\varepsilon_{\text{cor}}}\right)-2\log_2\left(\frac{2}{\varepsilon'\hat{\varepsilon}}\right)-2\log_2\left(\frac{1}{2\varepsilon_{\text{PA}}}\right).
    \end{equation}
To calculate the key rate, we need to estimate the lower bound of the count $\underline{Y}^Z_{111}$ and the upper bound of the phase error rate $\overline{e}^{PZ}_{111}$ for the single-photon component in the Z-basis. The specific methods are as follows \cite{lu2024repeaterlike}: 

\begin{itemize}
    \item {Estimation of statistical fluctuations}: Based on the approximate formula for the Chernoff bound given in the literature, given the expected value $\langle n\rangle$ and failure probability $\varepsilon$, the upper and lower bound estimates of the real value are respectively:
    \begin{equation}
        \overline{n}=\langle n\rangle+\frac{\beta}{2}+\sqrt{2\beta\langle n\rangle+\frac{\beta^2}{4}}, \quad \underline{n}=\max\left\{\langle n\rangle-\sqrt{2\beta\langle n\rangle},0\right\},
    \end{equation}
    where $\beta=-\ln{\varepsilon}$. Under the same conditions, given the observed value $n$, the upper and lower bounds of the expected value $\langle n\rangle$ are estimated as follows:
    \begin{equation}
        \overline{\langle n\rangle}=n+\beta+\sqrt{2\beta n+\beta^2} \quad 
        \underline{\langle n\rangle}=\max\left\{n-\frac{\beta}{2}-\sqrt{2\beta n+\frac{\beta^2}{4}}, 0 \right\}.
    \end{equation}

    \item {Pairing probability}: To calculate the expected values of pair count of Z-basis single-photon component and QBER of X-basis single-photon component, it is necessary to calculate the pulse emission probability $p_{[k_1^{\text{tot}},k_2^{\text{tot}},k_3^{\text{tot}}]}$ for different pairing patterns related to the total intensity of early and late pulses: 
    \begin{equation}
        p_{[k_{1}^{\text{tot}},k_{2}^{\text{tot}},...,k_{N}^{\text{tot}}]}=\frac{1}{p_{s}^{3}}\sum_{\substack{k_{i}^{e}+k_{i}^{l}=k_{i}^{\mathrm{tot}} \\
        \forall i\in\{1,...,N\}}
        }\prod_{i=1}^{N}p_{k_{i}^{e}}p_{k_{i}^{l}},
    \end{equation}
    where $N=3$ is the number of users, and $p_s=1-\sum_{k_1^ek_N^l,\dots, k_N^ek_{N-1}^l\in\{ \mu\nu,\nu\mu\}}q^{P_i}_{k_{i}^lk_{i+1}^e}$ is the correction factor for the filtering scheme, which can be eliminated in the calculation.

    \item {Pair count of Z-basis single-photon component}
    \begin{equation}
        \begin{aligned}
            \underline{\langle Y\rangle}_{111}^{Z} & \geq\frac{e^{-3\mu}p_{[\mu,\mu,\mu]}}{\nu^{3}(\mu-\nu)}\left[\mu^{4}\left(e^{3\nu}\frac{\underline{\langle n\rangle}_{[\nu,\nu,\nu]}}{p_{[\nu,\nu,\nu]}}-e^{2\nu}\frac{\overline{\langle n\rangle}_{[\nu,\nu,o]}}{p_{[\nu,\nu,o]}}-e^{2\nu}\frac{\overline{\langle n\rangle}_{[\nu,o,\nu]}}{p_{[\nu,o,\nu]}}-e^{2\nu}\frac{\overline{\langle n\rangle}_{[o,\nu,\nu]}}{p_{[o,\nu,\nu]}}+e^{\nu}\frac{\underline{\langle n\rangle}_{[\nu,o,o]}}{p_{[\nu,o,o]}}+e^{\nu}\frac{\underline{\langle n\rangle}_{[o,\nu,o]}}{p_{[o,\nu,o]}}\right.\right. \\
             & +e^{\nu}\frac{\underline{\langle n\rangle}_{[o,o,\nu]}}{p_{[o,o,\nu]}}-\frac{\overline{\langle n\rangle}_{[o,o,o]}}{p_{[o,o,o]}}\Bigg)-\nu^{4}\left(e^{3\mu}\frac{\overline{\langle n\rangle}_{[\mu,\mu,\mu]}}{p_{[\mu,\mu,\mu]}}-e^{2\mu}\frac{\underline{\langle n\rangle}_{[\mu,\mu,o]}}{p_{[\mu,\mu,o]}}-e^{2\mu}\frac{\underline{\langle n\rangle}_{[\mu,o,\mu]}}{p_{[\mu,o,\mu]}}-e^{2\mu}\frac{\underline{\langle n\rangle}_{[o,\mu,\mu]}}{p_{[o,\mu,\mu]}}\right. \\
             & +e^{\mu}\frac{\overline{\langle n\rangle}_{[\mu,o,o]}}{p_{[\mu,o,o]}}+e^{\mu}\frac{\overline{\langle n\rangle}_{[o,\mu,o]}}{p_{[o,\mu,o]}}+e^{\mu}\frac{\overline{\langle n\rangle}_{[o,o,\mu]}}{p_{[o,o,\mu]}}-\frac{\underline{\langle n\rangle}_{[o,o,o]}}{p_{[o,o,o]}}\Bigg)\Bigg].
        \end{aligned}
    \end{equation}

    \item {Pair count and QBER of X-basis single-photon component}
    
    The X-basis single-photon error pair count is:
    \begin{equation}
        \begin{aligned}
            \overline{\langle t\rangle}_{111}^{X} & \leq e^{-6\nu}p_{[2\nu,2\nu,2\nu]}\left(e^{6\nu}\frac{\overline{\langle m\rangle}_{[2\nu,2\nu,2\nu]}}{p_{[2\nu,2\nu,2\nu]}}-e^{4\nu}\frac{\underline{\langle n\rangle}_{[2\nu,2\nu,o]}}{2p_{[2\nu,2\nu,o]}}-e^{4\nu}\frac{\underline{\langle n\rangle}_{[2\nu,o,2\nu]}}{2p_{[2\nu,o,2\nu]}}-e^{4\nu}\frac{\underline{\langle n\rangle}_{[o,2\nu,2\nu]}}{2p_{[o,2\nu,2\nu]}}+e^{2\nu}\frac{\overline{\langle n\rangle}_{[2\nu,o,o]}}{2p_{[2\nu,o,o]}}\right. \\
             & +e^{2\nu}\frac{\overline{\langle n\rangle}_{[o,2\nu,o]}}{2p_{[o,2\nu,o]}}+e^{2\nu}\frac{\overline{\langle n\rangle}_{[o,o,2\nu]}}{2p_{[o,o,2\nu]}}-\frac{\underline{\langle n\rangle}_{[o,o,o]}}{2p_{[o,o,o]}}\bigg),
        \end{aligned}
    \end{equation}
    where ${\langle m\rangle}_{[2\nu,2\nu,2\nu]}$ represents the number of X-basis error pairing events. On the other hand, the expected value of the X-basis single-photon component pair count can be calculated  using the corresponding Z-basis count expected value $\underline{\langle Y\rangle}_{111}^{Z}$ calculated previously with proportional relation:
    \begin{equation}
        \frac{\underline{\langle Y\rangle}_{111}^{Z}}{\underline{\langle Y\rangle}_{111}^{X}}=\frac{\mu^3e^{-3\mu}p_{[\mu,\mu,\mu]}}{(2\nu e^{-2\nu})^3p_{[2\nu,2\nu,2\nu]}}.
    \end{equation}
    After estimating the bound of the true values from the above expected values, the upper bound of the X-basis single-photon QBER can be obtained as 
    $\overline{e}^X_{111}=\overline{t}^X_{111}/\underline{Y}^X_{111}$.

    \item {Random sampling formula and Z-basis single-photon phase error rate}: Under asymptotic conditions, the Z-basis single-photon phase error rate is identical to the X-basis single-photon bit error rate. However, under conditions of limited data volume, it is necessary to first estimate the corresponding true values based on the previously calculated expected values of each single-photon component, and then calculate $\overline{e}^{PZ}_{111}$ from these estimates. In the experiment, the random sampling theory used in the literature was adopted, i.e., the upper bound of the true value $\overline{\chi}=\lambda+\gamma^U(n,k,\lambda,\varepsilon)$, where
    \begin{equation}
        \gamma^U(n,k,\lambda,\varepsilon)=\left[\frac{(1-2\lambda)AG}{n+k}+\sqrt{\frac{A^2G^2}{(n+k)^2}+4\lambda(1-\lambda)G}\right]\left[2+2\frac{A^2G}{(n+k)^2}\right]^{-1},
    \end{equation}
    and
    \begin{equation}
        A=\max\{n,k\}, \quad G=\frac{n+k}{nk}\ln\frac{n+k}{2\pi nk\lambda\left(1-\lambda\right)\varepsilon^2}.
    \end{equation}
    Based on this, the upper bound of the true value of the Z-basis single-photon phase error rate can be finally obtained as:
    \begin{equation}
    \overline{e}^{PZ}_{111}=\overline{e}^X_{111}+\gamma\left(\underline{Y}^Z_{111},\underline{Y}^X_{111},\overline{e}^X_{111},\varepsilon\right).
    \end{equation}
\end{itemize}

\setcounter{section}{3}
\setcounter{equation}{0}
\section{S3 FFT-based frequency difference estimation}
{To estimate the frequency difference between users pairwise, in the experiment, we pair the single-count events on the three branches of the interferometer according to the method in MP QKD \cite{li_twin-field_2023,mp_zhang_experimental_2025,mp_shao_high-rate_2025,mp_lu_experimental_2025}, and calculate the X-basis QBER, which is given by the following formula:
\begin{equation}
		\begin{split}
			E_X=\frac{1}{2}-\frac{V_{2}}{2}\int_{-\infty}^{+\infty}\cos(2\pi\Delta f\Delta t+\omega\Delta t)G(\omega)\mathrm{d}\omega
			=\frac{1-V_{2}}{2}+\frac{V_{2}}{2}\left[1-e^{-\sigma^{2}\Delta t^{2}/2}\cos(2\pi\Delta f\Delta t)\right],
		\end{split}
	\end{equation}
in which the $V_2$ is visibility of HOM interference; $\Delta t$ is pairing time interval; $\omega$ is phase drift rate in fiber; and $G(\omega)$ is the Gaussian distribution. Since we did not incorporate a long optical fiber, $\omega$ can be omitted, i.e.,
\begin{equation}
			E_X\simeq\frac{1-V_{2}}{2}+\frac{V_{2}}{2}\left[1-\cos(2\pi\Delta f\Delta t)\right].
	\end{equation}
The $E_X(\Delta t)$ and their spectra obtained by FFT of the three pairs of users are shown in the following FIG. \ref{fig:s2}. Based on the protocol, the effect of frequency difference on the total relative phase is $\theta_{\Delta f}=\omega_A(t_1-t_3)+\omega_B(t_2-t_1)+\omega_C(t_3-t_2)$, which can be rewritten into a form related to the frequency difference between two users:
\begin{equation}
		\begin{split}
			\theta_{\Delta f}&=\omega_A(t_1-t_3)+\omega_B(t_2-t_1)+\omega_C(t_3-t_2)=(\omega_A-\omega_B)t_1+(\omega_B-\omega_C)t_2+(\omega_C-\omega_A)t_3\\&
			=(\omega_B-\omega_C)(t_2-t_1)+(\omega_C-\omega_A)(t_3-t_1)=\Delta\omega_{BC}\Delta t_{21}+\Delta\omega_{CA}\Delta t_{31}.
		\end{split}
\end{equation}
Similarly, we can get:
\begin{equation}
    \theta_{\Delta f}=\Delta\omega_{CB}\Delta t_{12}+\Delta\omega_{AC}\Delta t_{13}
        =\Delta\omega_{BA}\Delta t_{21}+\Delta\omega_{AC}\Delta t_{23}
        =\Delta\omega_{BA}\Delta t_{31}+\Delta\omega_{CB}\Delta t_{32}.
\end{equation}
After obtaining the frequency estimation values, we calculate the phase compensation values as:
\begin{equation}
    \theta_{\Delta f}=\frac{1}{3}\left[\left(\Delta\omega_{CB}\Delta t_{12}+\Delta\omega_{AC}\Delta t_{13}\right)+\left(\Delta\omega_{BA}\Delta t_{21}+\Delta\omega_{AC}\Delta t_{23}\right)
        +\left(\Delta\omega_{BA}\Delta t_{31}+\Delta\omega_{CB}\Delta t_{32}\right)\right].
\end{equation}
\begin{figure}[H]
	\centering
	\includegraphics[width=\linewidth]{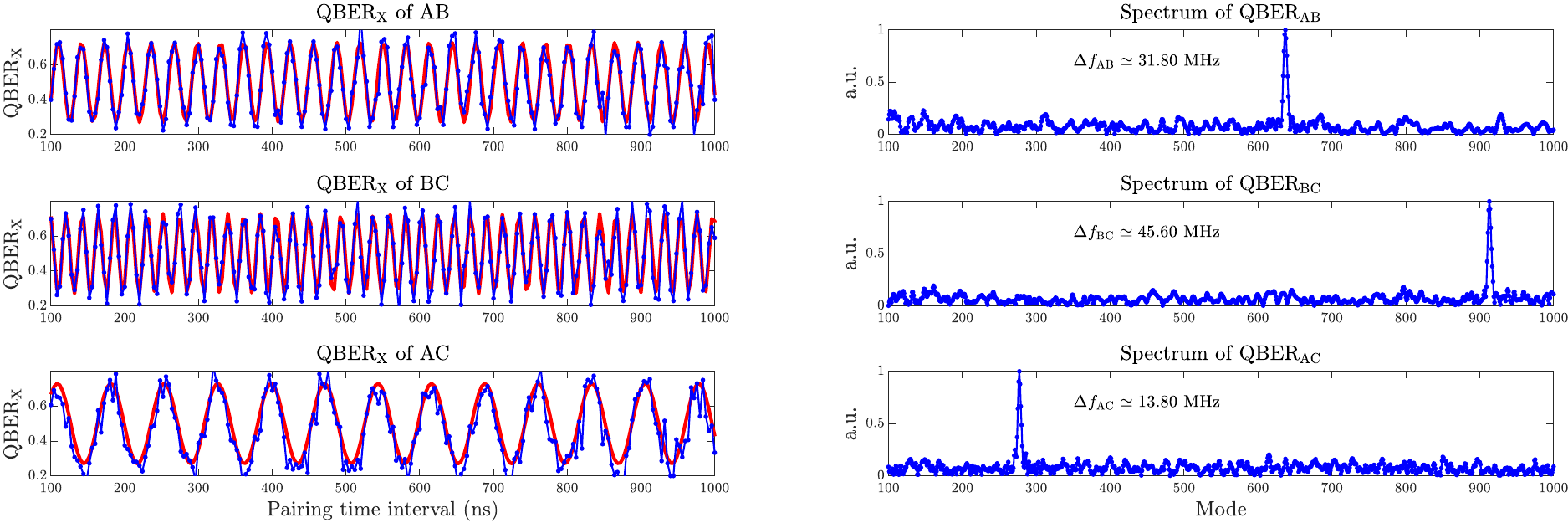}
	\caption{\label{fig:s2}Two-user X-basis QBERs of reference pulses in the time domain and their spectrum. (a) QBER versus pairing time interval. The blue experimental data points are in good agreement with the red fitting curve, whose HOM interference visibility parameter $V_2=0.46$ and frequency difference parameters are derived from the FFT calculation results. (b) Spectrum of QBERs with the FFT algorithm. 
    }
\end{figure}
When processing the data, the minimum pairing time interval is selected as the pulse period of 4 ns, and the FFT is performed on the data with pairing time intervals ranging from 0.1 $\mu$s to 20.1 $\mu$s, corresponding to a frequency difference estimation accuracy of about 50 kHz. When analyzing the QBER spectrum, we select the search region for the mode with the maximum amplitude according to the frequency difference measured at the initial moment, which is set to about 1 MHz–125 MHz in the experiment.
}

\section{S4 Experimental data and analysis}
\subsection{\label{sec:s4.1}S4.1 Parameters in key rate calculation and simulation}
The values of the common parameters used in data processing are as follows:

\setlength{\tabcolsep}{12pt}
\renewcommand\arraystretch{1.2}
\begin{table}[H]
\centering
\begin{tabular}{@{}ccccccccc@{}}
\toprule
$\mu$ & $\nu$ & $p_{\mu}$ & $p_{\nu}$ & $f$ & $\varepsilon$ & $ t_{\text{max}}$ & $e_d$ & $p_d$\\ \midrule
0.286 & 0.026 & 0.1008 & 0.5398 & 1.02  & $10^{-10}$  & 3 $\mu$s   & 0.01  & $4.2\times10^{-8}$                           \\ \bottomrule
\end{tabular}
\caption{\label{table:s1}Common parameters in data processing and simulation}
\end{table}

The public parameters used in the data processing and simulation are shown in Table \ref{table:s1} above. $p_{\mu}$ and $p_{\nu}$ represent the average probabilities of the user sending signal states and decoy states, respectively. The overall system loss is calculated based on the intensities and probabilities of the signal and decoy states, and the average single-photon counts of each detector. In the data processing, we filter the data with about 1 ns gate width, such that the dark count rate per pulse is about $4.2\times10^{-8} $. In the key rate calculation, we set the failure probability for Chernoff bound $\varepsilon$, the failure probability for error correction $\varepsilon_{\mathrm{cor}}$, the failure probability for privacy amplification $\varepsilon_{\mathrm{PA}}$ to $\varepsilon_{\mathrm{cor}}=\varepsilon_{\mathrm{PA}}=\varepsilon=10^{-10}$, and set the parameters of smooth entropies $\varepsilon'=\hat{\varepsilon}=10^{-7}$. Besides, more precisely, the data accumulation times under three different loss conditions are about 5003.90 seconds, 20006.85 seconds and 80074.01 seconds, respectively.

When simulating theoretical key rate curves under different experimental conditions, we set the Z-basis QBER to the corresponding measurement values as shown in Table \ref{table:s2}. On the other hand, based on the X-basis QBER data, we fit the GHZ-HOM interference visibility for the quantum pulses to be about $V=0.182$ at 39.3 dB loss, about $V=0.192$ at 48.6 dB, and about $V=0.191$ at 59.6 dB loss. Additionally, we set the phase drift rate $\omega$ in the long fiber to 0, and assume that the frequency difference $\Delta f=0$ between users after phase compensation.

\subsection{S4.2 Temporal structure of pulse sequence}
The schematic diagram of the pulse sequence prepared by the users in the experiment is presented in FIG. 1(c) of the main text. To illustrate the temporal structure of the pulse sequence more clearly and intuitively, the histograms of single-photon counts measured by a single detector are shown below, in which the width of each time bin, i.e., the time resolution of the TDC used in the experiment, is about 156.25 ps.

\begin{figure}[htb]
	\centering
	\includegraphics[width=\linewidth]{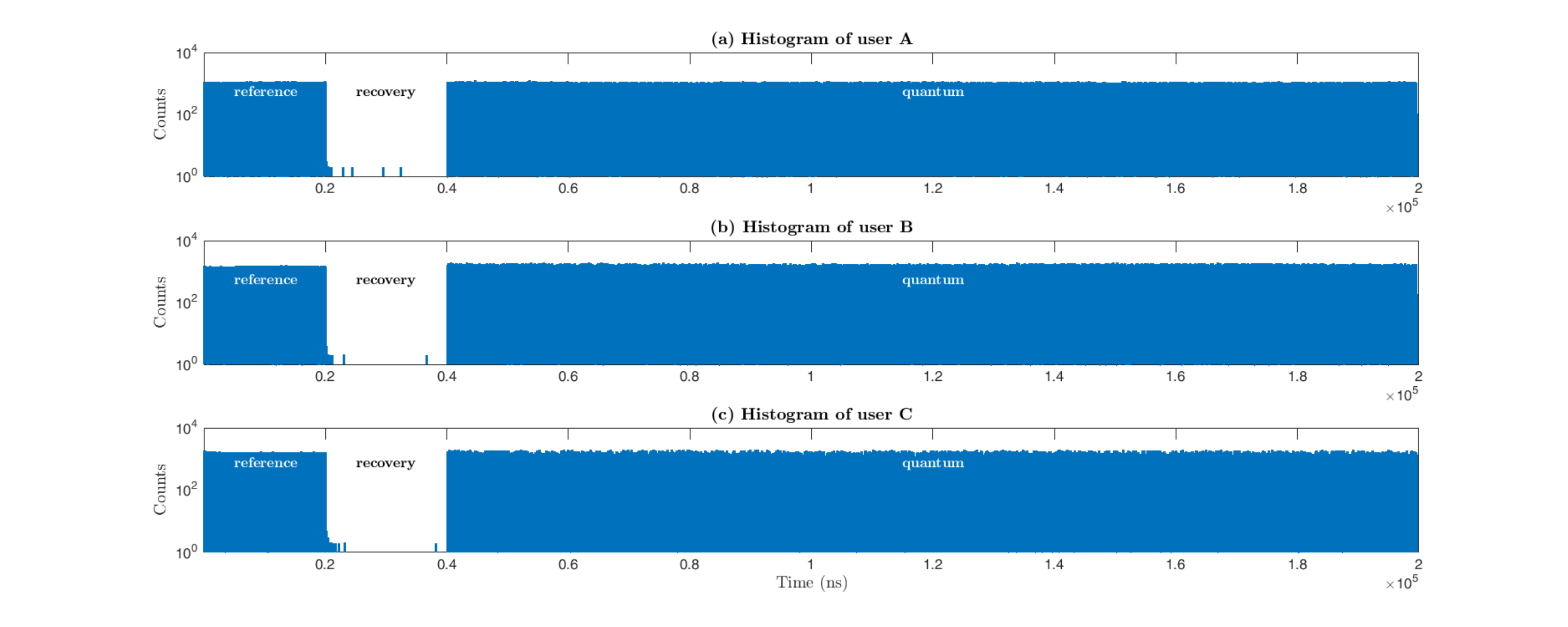}
    \caption{\label{fig:chap5_hist_all}Temporal histogram of the complete pulse sequence.}
\end{figure}
\begin{figure}[htb]
	\centering
	\includegraphics[width=\linewidth]{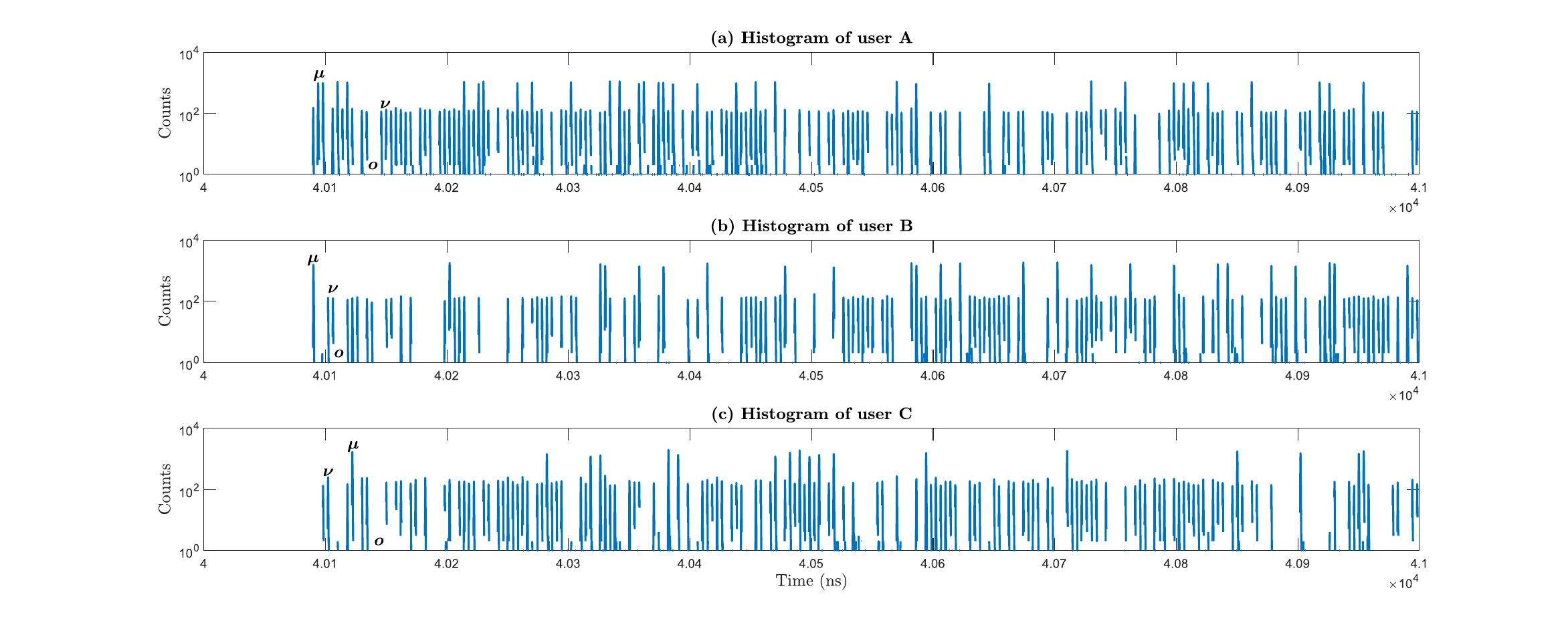}
    \caption{\label{fig:chap5_hist_qo}Temporal histogram of partial quantum pulses.}
\end{figure}

As shown in the figure above, the extinction ratio between the signal state and the vacuum state in the experiment exceeds 30 dB.
Besides, the average photon number of the reference pulses is approximately the same as that of the signal states of the quantum pulses. In the experiment, the reference pulse intensity is comprehensively determined based on factors including the number of two-photon and three-photon pairs, the X-basis QBER after frequency difference estimation and phase compensation, and the maximum count rate of the detectors.
Under the channel loss conditions selected in the experiment, the reference pulses with intensity close to that of the signal states can ensure excellent compensation effects for frequency difference and phase drift, while making the experimental adjustment process relatively simple.
Under higher channel loss conditions, the intensity of the reference pulses needs to be appropriately increased to ensure that sufficient pairs can be obtained within a relatively short time for accurate estimation and compensation of the frequency difference and phase drift.

\subsection{\label{sec:s4.3}S4.3 Distribution of pair counts by pairing time interval}
During data processing in the experiment, we select a maximum pairing time interval of 3 $\mu$s to reduce the X-basis QBER and increase the key rate. Under three different loss conditions, the pair counts exhibit similar distributions, with average pairing time intervals consistently about 2 $\mu$s. 
When the maximum pairing time interval is set to $t_{\mathrm{max}}=\infty_{Q}$, which means that single counts are paired without restriction on the pairing time interval within the quantum pulse section of the same pulse sequence shown in FIG. \ref{fig:scheme}(c) of the main text, the pairing probability exhibits a maximum point at about 10.35 $\mu$s, 13.72 $\mu$s and 20.64 $\mu$s, with average pairing time intervals of about 18.79 $\mu$s, 25.26 $\mu$s and 35.87 $\mu$s for 39.3 dB, 48.6 dB and 59.6 dB, respectively. In addition, the total number of pairs with $t_{\mathrm{max}}=3$ $\mu$s is approximately 64.39\%, 41.13\%, and 16.75\% of that with $t_{\mathrm{max}}=\infty$, i.e., no restriction imposed on the pairing time interval and pulse sequence where the single count is located, under the overall loss of about 39.3 dB, 48.6 dB and 59.6 dB, respectively. As the channel loss increases, the proportion of pairs under a finite pairing time interval gradually decreases, which is also one of the main difficulties to be overcome for realizing longer-distance communication in the future.

\begin{figure}[H]
	\centering
	\includegraphics[width=\linewidth]{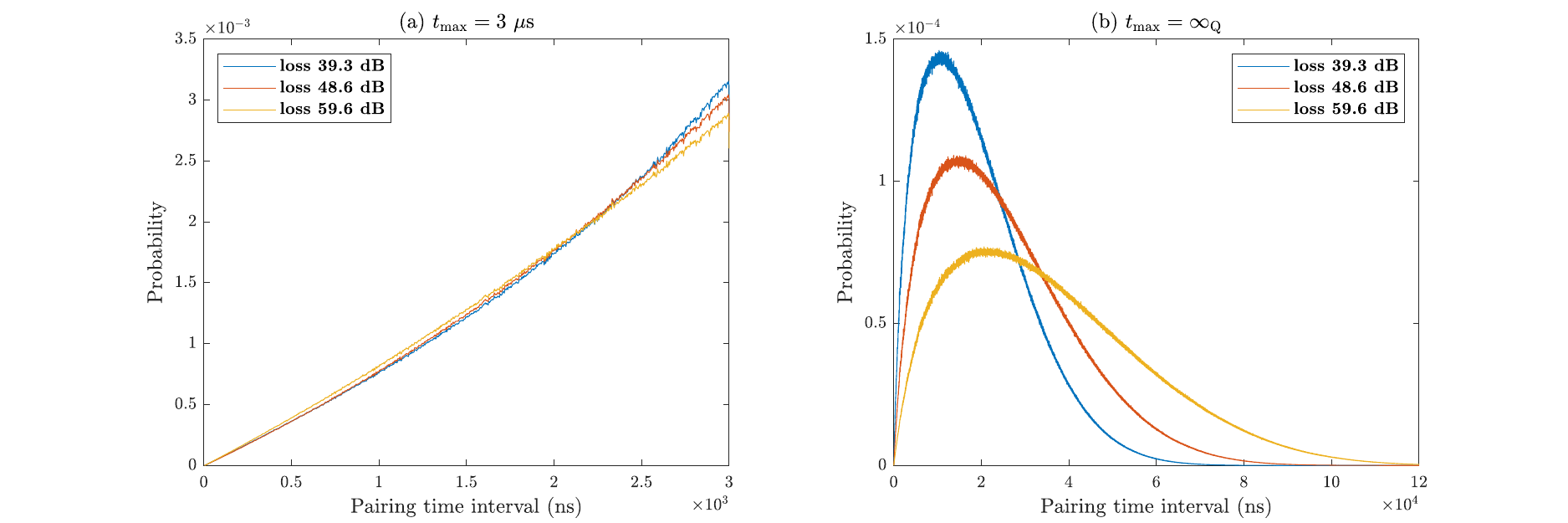}
	\caption{Distribution of pair counts by pairing time interval. (a) $t_{\mathrm{max}}=3$ $\mu$s, (b) $t_{\mathrm{max}}=\infty_Q$. 
    }
\end{figure}

\subsection{S4.4 Scale of total number of pairs and asymptotic key rate}
The advantage of the AMDI QCC protocol is that its key rate scales as $O(\eta)$ in the ideal case, which is independent of the number of users and can break the PLOB bound. In the experiment, due to the limitations of the total number of transmitted pulses, maximum pairing time interval, quantum bit error rates and other factors, we are unable to achieve communication performance breaking the PLOB bound.

To verify the linear relationship between the key rate and single-user channel transmittance, we calculate the total number of pairs per second for the maximum pairing time intervals $t_{\mathrm{max}}=3$ $\mu$s and $t_{\mathrm{max}}=\infty$ under different overall loss conditions in the experiment. As shown in FIG. \ref{fig:linear}(a) below, for case $t_{\mathrm{max}}=\infty$, we calculate the theoretical number of pairs using formula $n_{\mathrm{total}}=N\bar{\mu}\eta$ provided in the \cite{lu2024repeaterlike}, where $N$ is the total number of pulses transmitted by each user per second, $\bar\mu=p_{\mu}\mu+p_{\nu}\nu$ is the mean intensity of the pulses, and $\eta$ is the overall channel transmittance between a single user and the detection node. The calculation results show that the experimental results are in good agreement with theoretical expectations, thus demonstrating that the total number of pairs exhibits a linear relationship with channel transmittance in the case of $t_{\mathrm{max}}=\infty$.

On the other hand, as shown in FIG. \ref{fig:linear}(b) below, we calculate the asymptotic key rate from the experimental data without considering statistical fluctuations. Meanwhile, we proportionally calculate the asymptotic key rate expected to be obtained with $t_{\mathrm{max}}=\infty$ under experimental conditions, based on the ratio of the total number of pairs with $t_{\mathrm{max}}=3$ $\mu$s to that with $t_{\mathrm{max}}=\infty$ mentioned in Section \hyperref[sec:s4.3]{S4.3}. Then we calculate the theoretical key rate curves for $t_{\mathrm{max}}=3$ $\mu$s and $t_{\mathrm{max}}=\infty$ based on the methods given in \cite{lu2024repeaterlike}. Parameters such as the Z-basis QBER and the GHZ-HOM interference visibility are taken as the average values of the three sets of parameters presented in Section \hyperref[sec:s4.1]{S4.1}. For comparison, we take $10^{-4}\times \eta$ as the linear key rate reference, as indicated by the black curve in the figure. The asymptotic key rates under different scenarios are close to the theoretical values, while the data at the overall loss of 39.3 dB exhibits a lower key rate due to the relatively high error rate. Combined with the calculated results of the total number of pairs, the analysis of the asymptotic key rate shows that, under ideal conditions with a sufficiently large maximum pairing time interval, the key rate also exhibits a linear dependence on the single-user channel transmittance.

It is worth noting that the case of $t_{\mathrm{max}}=\infty$ analyzed above is an ideal scenario. In practice, on the one hand, when using independent lasers as the light sources, the frequency difference information between the sources cannot be accurately obtained, and the maximum allowable pairing time interval that can be selected in the experiment is also related to the precision of the frequency difference estimation. In the presence of estimation errors, although a larger pairing time interval provides more pairs, it simultaneously introduces larger phase estimation errors and a higher X‑basis QBER. On the other hand, in a practical long‑fiber link, the phase drift that accumulates over a longer pairing time interval within the long fiber link is non‑negligible and consequently also contributes to an increased X-basis QBER.

\begin{figure}[htb]
	\centering
	\includegraphics[width=\linewidth]{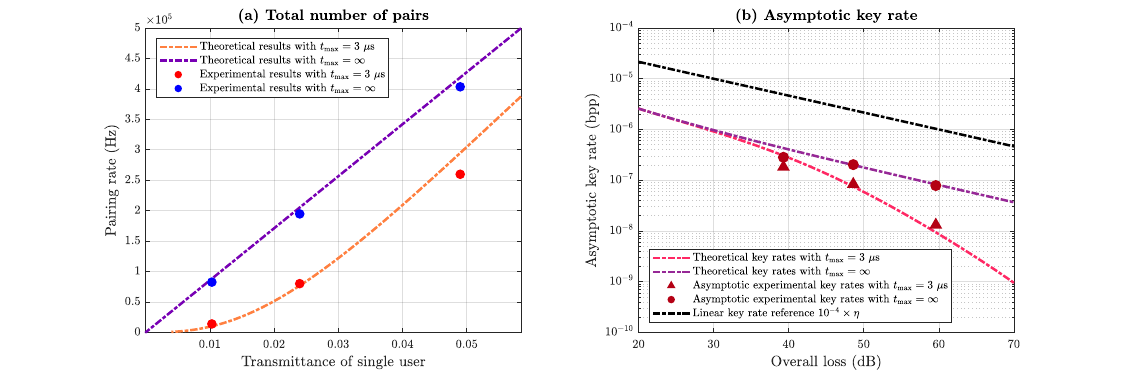}
    \caption{\label{fig:linear}Total number of pairs and asymptotic key rate. (a) Total number of pairs with $t_{\mathrm{max}}=3$ $\mu$s and $t_{\mathrm{max}}=\infty$ and corresponding theoretical results. (b) Asymptotic key rate in the experiment and theoretical simulation results with $t_{\mathrm{max}}=3$ $\mu$s and $t_{\mathrm{max}}=\infty$.}
\end{figure}
\subsection{S4.5 Long-term system stability}
The Z-basis QBER and X-basis QBER of the reference pulses with and without adopting frequency and phase drift compensation obtained during about 6 hours of testing are shown in the figure below. Experimental results indicate that the system can maintain stability over long-term data accumulation. 

\subsection{S4.6 Detailed experimental data}
The specific pairing count data under three different loss conditions are shown in the table below. The loss values listed in the table represent the overall total loss for the three users, including insertion loss and detection efficiency; $n_{[\alpha_1, \alpha_2, \alpha_3]}$ denotes the number of pairs whose the sum of the paired early and late pulse intensities for the three user pairs is $\alpha_1, \alpha_2, \alpha_3$, respectively; the superscripts $Z$ and $X$ respectively indicate the $Z$-basis and $X$-basis data after pairing and filtering. Especially, we take $n_{[\mu,\mu,\mu]}$ as $n_Z$ in the key rate formula \ref{eq:key_rate}. 
Besides, $\overline{\Delta t}$ represents the average pairing time interval, $L_{\text{min}}$ represents the minimum secure key length, and $N$ represents the total number of quantum pulses sent by users.

\begin{figure}[H]
	\centering
	\includegraphics[width=0.95\linewidth]{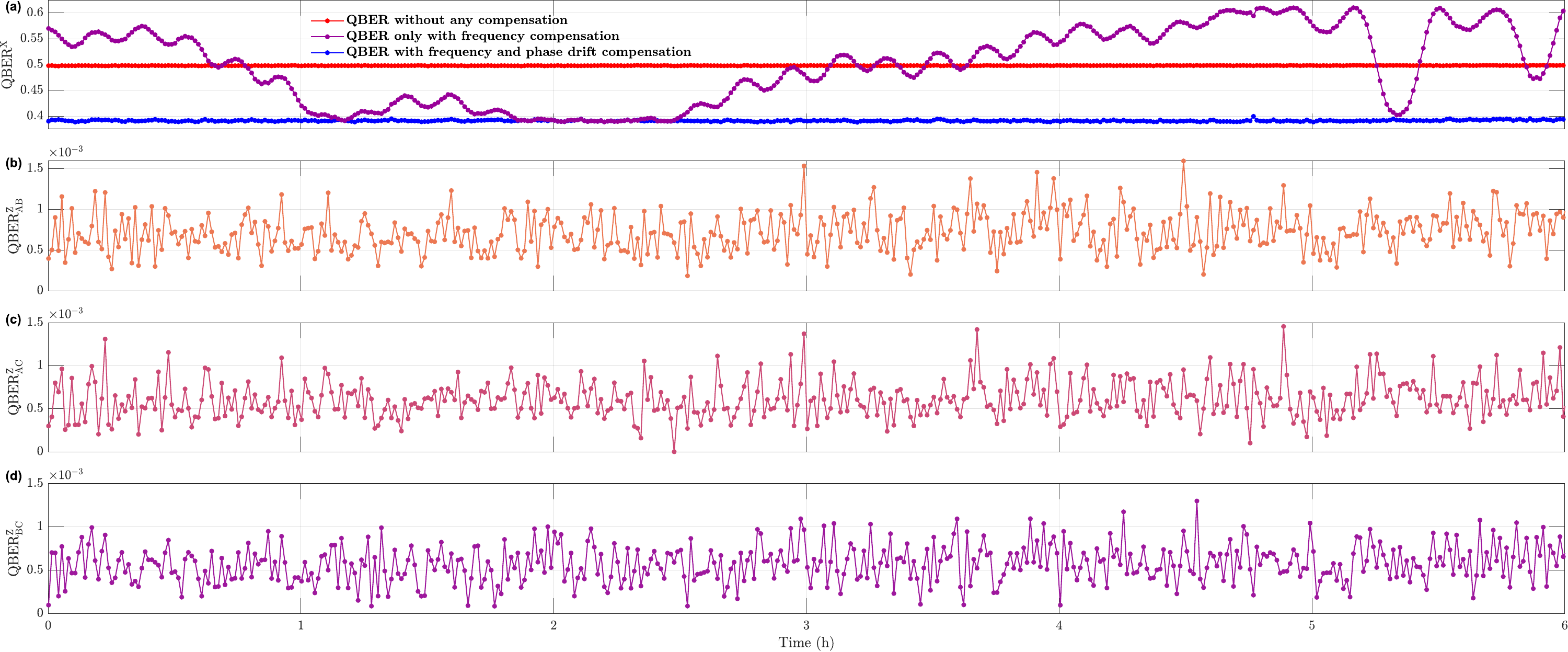}
	\caption{
     Stability of QBER. (a) X-basis QBER of reference pulses, (b) Z-basis QBER of users AB, (c) Z-basis QBER of users AC, (d) Z-basis QBER of users BC.
    }
\end{figure}

\setlength{\tabcolsep}{20pt}
\renewcommand\arraystretch{1.0}
\centering
\begin{longtable}{|c|l|lll|} 
\toprule
 & Att. (dB) & \textbf{39.3} & \textbf{48.6} & \textbf{59.6} \\* 
\hline
\textbf{Z-pair counts} & $n_{[\mu,\mu,\mu]}$ & 4536856& 5542901& 3864918\\* 
\hline
\multirow{33}{*}{\textbf{Total counts}} & $n_{[\nu,\nu,\nu]}$ & 585597 & 701399& 484697\\*
 & $n_{[o,o,o]}$ & 0 & 0 & 0 \\*
 & $n_{[\mu,o,o]}$ & 5 & 8 & 2 \\*
 & $n_{[o,\mu,o]}$ & 16 & 10 & 1 \\*
 & $n_{[o,o,\mu]}$ & 14 & 9 & 1 \\*
 & $n_{[\nu,o,o]}$ & 8 & 2 & 0 \\*
 & $n_{[o,\nu,o]}$ & 9 & 4 & 1 \\*
 & $n_{[o,o,\nu]}$ & 9 & 2 & 4 \\*
 & $n_{[o,\mu,\mu]}$ & 10408 & 7401 & 5012 \\*
 & $n_{[\mu,o,\mu]}$ & 11045 & 7962 & 4507 \\*
 & $n_{[\mu,\mu,o]}$ & 10527 & 6573 & 4659 \\*
 & $n_{[o,\nu,\nu]}$ & 2620 & 1724  & 1169 \\*
 & $n_{[\nu,o,\nu]}$ & 2827 & 1947  & 1129 \\*
 & $n_{[\nu,\nu,o]}$ & 2564 & 1590  & 1114 \\*
 & $n_{[\mu,\nu,\nu]}$ & 1168124& 1391282& 960283\\*
 & $n_{[\nu,\mu,\nu]}$ & 1142388& 1363918& 966002\\*
 & $n_{[\nu,\nu,\mu]}$ & 1174619& 1447332& 983941\\*
 & $n_{[\nu,\mu,\mu]}$ & 2275619& 2791992& 1946789\\*
 & $n_{[\mu,\nu,\mu]}$ & 2349050& 2886105& 1958907\\*
 & $n_{[\mu,\mu,\nu]}$ & 2268010& 2698297& 1907766\\*
 & $n_{[\mu,\nu,o]}$ & 5166 & 3380 & 2290 \\*
 & $n_{[\mu,o,\nu]}$ & 5584 & 3918 & 2266 \\*
 & $n_{[\nu,\mu,o]}$ & 5267 & 3370 & 2382 \\*
 & $n_{[\nu,o,\mu]}$ & 5507 & 4022 & 2301 \\*
 & $n_{[o,\mu,\nu]}$ & 5073 & 3529 & 2454 \\*
 & $n_{[o,\nu,\mu]}$ & 5325 & 3801 & 2495 \\*
 & $n_{[2\nu,2\nu,2\nu]}$ & 7740134& 9509079& 6553982\\*
 & $n_{[2\nu,o,o]}$ & 603 & 339 & 172  \\*
 & $n_{[o,2\nu,o]}$ & 944 & 707 & 170 \\*
 & $n_{[o,o,2\nu]}$ & 991 & 741 & 870 \\*
 & $n_{[o,2\nu,2\nu]}$ & 841063& 1058942& 737073\\*
 & $n_{[2\nu,o,2\nu]}$ & 883452& 1172157& 716824\\*
 & $n_{[2\nu,2\nu,o]}$ & 842577& 922802& 716352\\ 
\hline
\textbf{X-pair counts} & $n^X_{[2\nu,2\nu,2\nu]}$ & 967876& 1189572& 820887\\* 
\hline
\multirow{4}{*}{\textbf{Error counts}} & $E^{Z_{AB}}_{[\mu,\mu,\mu]}$ & 7778 & 5242 & 4353 \\*
 & $E^{Z_{AC}}_{[\mu,\mu,\mu]}$ & 7678 & 5400 & 2769 \\*
 & $E^{Z_{BC}}_{[\mu,\mu,\mu]}$ & 6693 & 4335 & 2805 \\*
 & $E^X_{[2\nu,2\nu,2\nu]}$ & 399992 & 486164 & 335669 \\* 
\hline
\multirow{8}{*}{\textbf{Results}} 
 & $\overline{\Delta t}$ ($\mu$s) & 2.076 & 2.067 & 2.042 \\*
 & $e^Z_{AB}$ (\textperthousand) & 1.714 & 0.946 & 1.126 \\*
 & $e^Z_{AC}$ (\textperthousand) & 1.692 & 0.974 & 0.716 \\*
 & $e^Z_{BC}$ (\textperthousand) & 1.475 & 0.782 & 0.726 \\*
 & $\underline{Y}^Z_{111}$ & $1.951\times 10^6$ & $2.391\times 10^6$ & $1.643\times 10^6$ \\*
 & $\overline{e}^{PZ}_{111}$ & 0.3528 & 0.3232 & 0.3413 \\*
 & $L_{\text{min}}$ & $3.932\times 10^{4}$ & $1.571\times 10^{5}$ & $7.139\times 10^{4}$ \\*
 & $L_{\text{min}}/N$ & $3.940\times 10^{-8}$ & $3.937\times 10^{-8}$ & $4.470\times 10^{-9}$ \\
\bottomrule
\caption{\label{table:s2}Detailed results of the AMDI QCC experiment}
\end{longtable}

\nocite{1}
\bibliography{amdi-qcc-bib}

\begin{thebibliography}{68}%
\makeatletter
\providecommand \@ifxundefined [1]{%
 \@ifx{#1\undefined}
}%
\providecommand \@ifnum [1]{%
 \ifnum #1\expandafter \@firstoftwo
 \else \expandafter \@secondoftwo
 \fi
}%
\providecommand \@ifx [1]{%
 \ifx #1\expandafter \@firstoftwo
 \else \expandafter \@secondoftwo
 \fi
}%
\providecommand \natexlab [1]{#1}%
\providecommand \enquote  [1]{``#1''}%
\providecommand \bibnamefont  [1]{#1}%
\providecommand \bibfnamefont [1]{#1}%
\providecommand \citenamefont [1]{#1}%
\providecommand \href@noop [0]{\@secondoftwo}%
\providecommand \href [0]{\begingroup \@sanitize@url \@href}%
\providecommand \@href[1]{\@@startlink{#1}\@@href}%
\providecommand \@@href[1]{\endgroup#1\@@endlink}%
\providecommand \@sanitize@url [0]{\catcode `\\12\catcode `\$12\catcode
  `\&12\catcode `\#12\catcode `\^12\catcode `\_12\catcode `\%12\relax}%
\providecommand \@@startlink[1]{}%
\providecommand \@@endlink[0]{}%
\providecommand \url  [0]{\begingroup\@sanitize@url \@url }%
\providecommand \@url [1]{\endgroup\@href {#1}{\urlprefix }}%
\providecommand \urlprefix  [0]{URL }%
\providecommand \Eprint [0]{\href }%
\providecommand \doibase [0]{https://doi.org/}%
\providecommand \selectlanguage [0]{\@gobble}%
\providecommand \bibinfo  [0]{\@secondoftwo}%
\providecommand \bibfield  [0]{\@secondoftwo}%
\providecommand \translation [1]{[#1]}%
\providecommand \BibitemOpen [0]{}%
\providecommand \bibitemStop [0]{}%
\providecommand \bibitemNoStop [0]{.\EOS\space}%
\providecommand \EOS [0]{\spacefactor3000\relax}%
\providecommand \BibitemShut  [1]{\csname bibitem#1\endcsname}%
\let\auto@bib@innerbib\@empty
\bibitem [{\citenamefont {Wehner}\ \emph {et~al.}(2018)\citenamefont {Wehner},
  \citenamefont {Elkouss},\ and\ \citenamefont {Hanson}}]{qnet_review_1}%
  \BibitemOpen
  \bibfield  {author} {\bibinfo {author} {\bibfnamefont {S.}~\bibnamefont
  {Wehner}}, \bibinfo {author} {\bibfnamefont {D.}~\bibnamefont {Elkouss}},\
  and\ \bibinfo {author} {\bibfnamefont {R.}~\bibnamefont {Hanson}},\
  }\bibfield  {title} {\bibinfo {title} {Quantum internet: A vision for the
  road ahead},\ }\href {https://doi.org/10.1126/science.aam9288} {\bibfield
  {journal} {\bibinfo  {journal} {Science}\ }\textbf {\bibinfo {volume}
  {362}},\ \bibinfo {pages} {eaam9288} (\bibinfo {year} {2018})}\BibitemShut
  {NoStop}%
\bibitem [{\citenamefont {Kimble}(2008)}]{qnet_review_2}%
  \BibitemOpen
  \bibfield  {author} {\bibinfo {author} {\bibfnamefont {H.~J.}\ \bibnamefont
  {Kimble}},\ }\bibfield  {title} {\bibinfo {title} {The quantum internet},\
  }\href {https://doi.org/10.1038/nature07127} {\bibfield  {journal} {\bibinfo
  {journal} {Nature}\ }\textbf {\bibinfo {volume} {453}},\ \bibinfo {pages}
  {1023} (\bibinfo {year} {2008})}\BibitemShut {NoStop}%
\bibitem [{\citenamefont {Bennett}\ and\ \citenamefont
  {Brassard}(1984)}]{BB84}%
  \BibitemOpen
  \bibfield  {author} {\bibinfo {author} {\bibfnamefont {C.~H.}\ \bibnamefont
  {Bennett}}\ and\ \bibinfo {author} {\bibfnamefont {G.}~\bibnamefont
  {Brassard}},\ }\bibinfo {title} {Quantum cryptography: Public key
  distribution and coin tossing},\ in\ \href@noop {} {\emph {\bibinfo
  {booktitle} {Proceedings of IEEE International Conference on Computers,
  Systems, and Signal Processing}}}\ (\bibinfo  {publisher} {IEEE},\ \bibinfo
  {address} {New York},\ \bibinfo {year} {1984})\ p.\ \bibinfo {pages}
  {175–179}\BibitemShut {NoStop}%
\bibitem [{\citenamefont {Ekert}(1991)}]{qkd_review_4}%
  \BibitemOpen
  \bibfield  {author} {\bibinfo {author} {\bibfnamefont {A.~K.}\ \bibnamefont
  {Ekert}},\ }\bibfield  {title} {\bibinfo {title} {Quantum cryptography based
  on bell's theorem},\ }\href {https://doi.org/10.1103/PhysRevLett.67.661}
  {\bibfield  {journal} {\bibinfo  {journal} {Phys. Rev. Lett.}\ }\textbf
  {\bibinfo {volume} {67}},\ \bibinfo {pages} {661} (\bibinfo {year}
  {1991})}\BibitemShut {NoStop}%
\bibitem [{\citenamefont {Gisin}\ \emph {et~al.}(2002)\citenamefont {Gisin},
  \citenamefont {Ribordy}, \citenamefont {Tittel},\ and\ \citenamefont
  {Zbinden}}]{qkd_review_1}%
  \BibitemOpen
  \bibfield  {author} {\bibinfo {author} {\bibfnamefont {N.}~\bibnamefont
  {Gisin}}, \bibinfo {author} {\bibfnamefont {G.}~\bibnamefont {Ribordy}},
  \bibinfo {author} {\bibfnamefont {W.}~\bibnamefont {Tittel}},\ and\ \bibinfo
  {author} {\bibfnamefont {H.}~\bibnamefont {Zbinden}},\ }\bibfield  {title}
  {\bibinfo {title} {Quantum cryptography},\ }\href
  {https://doi.org/10.1103/RevModPhys.74.145} {\bibfield  {journal} {\bibinfo
  {journal} {Rev. Mod. Phys.}\ }\textbf {\bibinfo {volume} {74}},\ \bibinfo
  {pages} {145} (\bibinfo {year} {2002})}\BibitemShut {NoStop}%
\bibitem [{\citenamefont {Scarani}\ \emph {et~al.}(2009)\citenamefont
  {Scarani}, \citenamefont {Bechmann-Pasquinucci}, \citenamefont {Cerf},
  \citenamefont {Du\ifmmode~\check{s}\else \v{s}\fi{}ek}, \citenamefont
  {L\"utkenhaus},\ and\ \citenamefont {Peev}}]{qkd_review_2}%
  \BibitemOpen
  \bibfield  {author} {\bibinfo {author} {\bibfnamefont {V.}~\bibnamefont
  {Scarani}}, \bibinfo {author} {\bibfnamefont {H.}~\bibnamefont
  {Bechmann-Pasquinucci}}, \bibinfo {author} {\bibfnamefont {N.~J.}\
  \bibnamefont {Cerf}}, \bibinfo {author} {\bibfnamefont {M.}~\bibnamefont
  {Du\ifmmode~\check{s}\else \v{s}\fi{}ek}}, \bibinfo {author} {\bibfnamefont
  {N.}~\bibnamefont {L\"utkenhaus}},\ and\ \bibinfo {author} {\bibfnamefont
  {M.}~\bibnamefont {Peev}},\ }\bibfield  {title} {\bibinfo {title} {The
  security of practical quantum key distribution},\ }\href
  {https://doi.org/10.1103/RevModPhys.81.1301} {\bibfield  {journal} {\bibinfo
  {journal} {Rev. Mod. Phys.}\ }\textbf {\bibinfo {volume} {81}},\ \bibinfo
  {pages} {1301} (\bibinfo {year} {2009})}\BibitemShut {NoStop}%
\bibitem [{\citenamefont {Xu}\ \emph {et~al.}(2020)\citenamefont {Xu},
  \citenamefont {Ma}, \citenamefont {Zhang}, \citenamefont {Lo},\ and\
  \citenamefont {Pan}}]{qkd_review_3}%
  \BibitemOpen
  \bibfield  {author} {\bibinfo {author} {\bibfnamefont {F.}~\bibnamefont
  {Xu}}, \bibinfo {author} {\bibfnamefont {X.}~\bibnamefont {Ma}}, \bibinfo
  {author} {\bibfnamefont {Q.}~\bibnamefont {Zhang}}, \bibinfo {author}
  {\bibfnamefont {H.-K.}\ \bibnamefont {Lo}},\ and\ \bibinfo {author}
  {\bibfnamefont {J.-W.}\ \bibnamefont {Pan}},\ }\bibfield  {title} {\bibinfo
  {title} {Secure quantum key distribution with realistic devices},\ }\href
  {https://doi.org/10.1103/RevModPhys.92.025002} {\bibfield  {journal}
  {\bibinfo  {journal} {Rev. Mod. Phys.}\ }\textbf {\bibinfo {volume} {92}},\
  \bibinfo {pages} {025002} (\bibinfo {year} {2020})}\BibitemShut {NoStop}%
\bibitem [{\citenamefont {Pirandola}\ \emph {et~al.}(2020)\citenamefont
  {Pirandola}, \citenamefont {Andersen}, \citenamefont {Banchi}, \citenamefont
  {Berta}, \citenamefont {Bunandar}, \citenamefont {Colbeck}, \citenamefont
  {Englund}, \citenamefont {Gehring}, \citenamefont {Lupo}, \citenamefont
  {Ottaviani}, \citenamefont {Pereira}, \citenamefont {Razavi}, \citenamefont
  {Shaari}, \citenamefont {Tomamichel}, \citenamefont {Usenko}, \citenamefont
  {Vallone}, \citenamefont {Villoresi},\ and\ \citenamefont
  {Wallden}}]{qkd_review_5}%
  \BibitemOpen
  \bibfield  {author} {\bibinfo {author} {\bibfnamefont {S.}~\bibnamefont
  {Pirandola}}, \bibinfo {author} {\bibfnamefont {U.~L.}\ \bibnamefont
  {Andersen}}, \bibinfo {author} {\bibfnamefont {L.}~\bibnamefont {Banchi}},
  \bibinfo {author} {\bibfnamefont {M.}~\bibnamefont {Berta}}, \bibinfo
  {author} {\bibfnamefont {D.}~\bibnamefont {Bunandar}}, \bibinfo {author}
  {\bibfnamefont {R.}~\bibnamefont {Colbeck}}, \bibinfo {author} {\bibfnamefont
  {D.}~\bibnamefont {Englund}}, \bibinfo {author} {\bibfnamefont
  {T.}~\bibnamefont {Gehring}}, \bibinfo {author} {\bibfnamefont
  {C.}~\bibnamefont {Lupo}}, \bibinfo {author} {\bibfnamefont {C.}~\bibnamefont
  {Ottaviani}}, \bibinfo {author} {\bibfnamefont {J.~L.}\ \bibnamefont
  {Pereira}}, \bibinfo {author} {\bibfnamefont {M.}~\bibnamefont {Razavi}},
  \bibinfo {author} {\bibfnamefont {J.~S.}\ \bibnamefont {Shaari}}, \bibinfo
  {author} {\bibfnamefont {M.}~\bibnamefont {Tomamichel}}, \bibinfo {author}
  {\bibfnamefont {V.~C.}\ \bibnamefont {Usenko}}, \bibinfo {author}
  {\bibfnamefont {G.}~\bibnamefont {Vallone}}, \bibinfo {author} {\bibfnamefont
  {P.}~\bibnamefont {Villoresi}},\ and\ \bibinfo {author} {\bibfnamefont
  {P.}~\bibnamefont {Wallden}},\ }\bibfield  {title} {\bibinfo {title}
  {Advances in quantum cryptography},\ }\href
  {https://doi.org/10.1364/AOP.361502} {\bibfield  {journal} {\bibinfo
  {journal} {Adv. Opt. Photon.}\ }\textbf {\bibinfo {volume} {12}},\ \bibinfo
  {pages} {1012} (\bibinfo {year} {2020})}\BibitemShut {NoStop}%
\bibitem [{\citenamefont {Hwang}(2003)}]{decoy1}%
  \BibitemOpen
  \bibfield  {author} {\bibinfo {author} {\bibfnamefont {W.-Y.}\ \bibnamefont
  {Hwang}},\ }\bibfield  {title} {\bibinfo {title} {Quantum key distribution
  with high loss: Toward global secure communication},\ }\href
  {https://doi.org/10.1103/PhysRevLett.91.057901} {\bibfield  {journal}
  {\bibinfo  {journal} {Phys. Rev. Lett.}\ }\textbf {\bibinfo {volume} {91}},\
  \bibinfo {pages} {057901} (\bibinfo {year} {2003})}\BibitemShut {NoStop}%
\bibitem [{\citenamefont {Wang}(2005)}]{decoy2}%
  \BibitemOpen
  \bibfield  {author} {\bibinfo {author} {\bibfnamefont {X.-B.}\ \bibnamefont
  {Wang}},\ }\bibfield  {title} {\bibinfo {title} {Beating the
  photon-number-splitting attack in practical quantum cryptography},\ }\href
  {https://doi.org/10.1103/PhysRevLett.94.230503} {\bibfield  {journal}
  {\bibinfo  {journal} {Phys. Rev. Lett.}\ }\textbf {\bibinfo {volume} {94}},\
  \bibinfo {pages} {230503} (\bibinfo {year} {2005})}\BibitemShut {NoStop}%
\bibitem [{\citenamefont {Lo}\ \emph {et~al.}(2005)\citenamefont {Lo},
  \citenamefont {Ma},\ and\ \citenamefont {Chen}}]{decoy3}%
  \BibitemOpen
  \bibfield  {author} {\bibinfo {author} {\bibfnamefont {H.-K.}\ \bibnamefont
  {Lo}}, \bibinfo {author} {\bibfnamefont {X.}~\bibnamefont {Ma}},\ and\
  \bibinfo {author} {\bibfnamefont {K.}~\bibnamefont {Chen}},\ }\bibfield
  {title} {\bibinfo {title} {Decoy state quantum key distribution},\ }\href
  {https://doi.org/10.1103/PhysRevLett.94.230504} {\bibfield  {journal}
  {\bibinfo  {journal} {Phys. Rev. Lett.}\ }\textbf {\bibinfo {volume} {94}},\
  \bibinfo {pages} {230504} (\bibinfo {year} {2005})}\BibitemShut {NoStop}%
\bibitem [{\citenamefont {Ac\'{\i}n}\ \emph {et~al.}(2007)\citenamefont
  {Ac\'{\i}n}, \citenamefont {Brunner}, \citenamefont {Gisin}, \citenamefont
  {Massar}, \citenamefont {Pironio},\ and\ \citenamefont {Scarani}}]{diqkd}%
  \BibitemOpen
  \bibfield  {author} {\bibinfo {author} {\bibfnamefont {A.}~\bibnamefont
  {Ac\'{\i}n}}, \bibinfo {author} {\bibfnamefont {N.}~\bibnamefont {Brunner}},
  \bibinfo {author} {\bibfnamefont {N.}~\bibnamefont {Gisin}}, \bibinfo
  {author} {\bibfnamefont {S.}~\bibnamefont {Massar}}, \bibinfo {author}
  {\bibfnamefont {S.}~\bibnamefont {Pironio}},\ and\ \bibinfo {author}
  {\bibfnamefont {V.}~\bibnamefont {Scarani}},\ }\bibfield  {title} {\bibinfo
  {title} {Device-independent security of quantum cryptography against
  collective attacks},\ }\href {https://doi.org/10.1103/PhysRevLett.98.230501}
  {\bibfield  {journal} {\bibinfo  {journal} {Phys. Rev. Lett.}\ }\textbf
  {\bibinfo {volume} {98}},\ \bibinfo {pages} {230501} (\bibinfo {year}
  {2007})}\BibitemShut {NoStop}%
\bibitem [{\citenamefont {Braunstein}\ and\ \citenamefont
  {Pirandola}(2012)}]{Pirandola2012}%
  \BibitemOpen
  \bibfield  {author} {\bibinfo {author} {\bibfnamefont {S.~L.}\ \bibnamefont
  {Braunstein}}\ and\ \bibinfo {author} {\bibfnamefont {S.}~\bibnamefont
  {Pirandola}},\ }\bibfield  {title} {\bibinfo {title} {Side-channel-free
  quantum key distribution},\ }\href
  {https://doi.org/10.1103/PhysRevLett.108.130502} {\bibfield  {journal}
  {\bibinfo  {journal} {Phys. Rev. Lett.}\ }\textbf {\bibinfo {volume} {108}},\
  \bibinfo {pages} {130502} (\bibinfo {year} {2012})}\BibitemShut {NoStop}%
\bibitem [{\citenamefont {Lo}\ \emph {et~al.}(2012)\citenamefont {Lo},
  \citenamefont {Curty},\ and\ \citenamefont {Qi}}]{mdi_qkd}%
  \BibitemOpen
  \bibfield  {author} {\bibinfo {author} {\bibfnamefont {H.-K.}\ \bibnamefont
  {Lo}}, \bibinfo {author} {\bibfnamefont {M.}~\bibnamefont {Curty}},\ and\
  \bibinfo {author} {\bibfnamefont {B.}~\bibnamefont {Qi}},\ }\bibfield
  {title} {\bibinfo {title} {Measurement-device-independent quantum key
  distribution},\ }\href {https://doi.org/10.1103/PhysRevLett.108.130503}
  {\bibfield  {journal} {\bibinfo  {journal} {Phys. Rev. Lett.}\ }\textbf
  {\bibinfo {volume} {108}},\ \bibinfo {pages} {130503} (\bibinfo {year}
  {2012})}\BibitemShut {NoStop}%
\bibitem [{\citenamefont {Lucamarini}\ \emph {et~al.}(2018)\citenamefont
  {Lucamarini}, \citenamefont {Yuan}, \citenamefont {Dynes},\ and\
  \citenamefont {Shields}}]{tf_qkd}%
  \BibitemOpen
  \bibfield  {author} {\bibinfo {author} {\bibfnamefont {M.}~\bibnamefont
  {Lucamarini}}, \bibinfo {author} {\bibfnamefont {Z.~L.}\ \bibnamefont
  {Yuan}}, \bibinfo {author} {\bibfnamefont {J.~F.}\ \bibnamefont {Dynes}},\
  and\ \bibinfo {author} {\bibfnamefont {A.~J.}\ \bibnamefont {Shields}},\
  }\bibfield  {title} {\bibinfo {title} {Overcoming the rate–distance limit
  of quantum key distribution without quantum repeaters},\ }\href
  {https://doi.org/10.1038/s41586-018-0066-6} {\bibfield  {journal} {\bibinfo
  {journal} {Nature}\ }\textbf {\bibinfo {volume} {557}},\ \bibinfo {pages}
  {400} (\bibinfo {year} {2018})}\BibitemShut {NoStop}%
\bibitem [{\citenamefont {Xie}\ \emph {et~al.}(2022)\citenamefont {Xie},
  \citenamefont {Lu}, \citenamefont {Weng}, \citenamefont {Cao}, \citenamefont
  {Jia}, \citenamefont {Bao}, \citenamefont {Wang}, \citenamefont {Fu},
  \citenamefont {Yin},\ and\ \citenamefont {Chen}}]{mp_qkd_2}%
  \BibitemOpen
  \bibfield  {author} {\bibinfo {author} {\bibfnamefont {Y.-M.}\ \bibnamefont
  {Xie}}, \bibinfo {author} {\bibfnamefont {Y.-S.}\ \bibnamefont {Lu}},
  \bibinfo {author} {\bibfnamefont {C.-X.}\ \bibnamefont {Weng}}, \bibinfo
  {author} {\bibfnamefont {X.-Y.}\ \bibnamefont {Cao}}, \bibinfo {author}
  {\bibfnamefont {Z.-Y.}\ \bibnamefont {Jia}}, \bibinfo {author} {\bibfnamefont
  {Y.}~\bibnamefont {Bao}}, \bibinfo {author} {\bibfnamefont {Y.}~\bibnamefont
  {Wang}}, \bibinfo {author} {\bibfnamefont {Y.}~\bibnamefont {Fu}}, \bibinfo
  {author} {\bibfnamefont {H.-L.}\ \bibnamefont {Yin}},\ and\ \bibinfo {author}
  {\bibfnamefont {Z.-B.}\ \bibnamefont {Chen}},\ }\bibfield  {title} {\bibinfo
  {title} {Breaking the rate-loss bound of quantum key distribution with
  asynchronous two-photon interference},\ }\href
  {https://doi.org/10.1103/PRXQuantum.3.020315} {\bibfield  {journal} {\bibinfo
   {journal} {PRX Quantum}\ }\textbf {\bibinfo {volume} {3}},\ \bibinfo {pages}
  {020315} (\bibinfo {year} {2022})}\BibitemShut {NoStop}%
\bibitem [{\citenamefont {Zeng}\ \emph {et~al.}(2022)\citenamefont {Zeng},
  \citenamefont {Zhou}, \citenamefont {Wu},\ and\ \citenamefont
  {Ma}}]{mp_qkd_1}%
  \BibitemOpen
  \bibfield  {author} {\bibinfo {author} {\bibfnamefont {P.}~\bibnamefont
  {Zeng}}, \bibinfo {author} {\bibfnamefont {H.}~\bibnamefont {Zhou}}, \bibinfo
  {author} {\bibfnamefont {W.}~\bibnamefont {Wu}},\ and\ \bibinfo {author}
  {\bibfnamefont {X.}~\bibnamefont {Ma}},\ }\bibfield  {title} {\bibinfo
  {title} {Mode-pairing quantum key distribution},\ }\href
  {https://doi.org/10.1038/s41467-022-31534-7} {\bibfield  {journal} {\bibinfo
  {journal} {Nature Communications}\ }\textbf {\bibinfo {volume} {13}},\
  \bibinfo {pages} {3903} (\bibinfo {year} {2022})}\BibitemShut {NoStop}%
\bibitem [{\citenamefont {Yin}\ \emph {et~al.}(2020)\citenamefont {Yin},
  \citenamefont {Li}, \citenamefont {Liao}, \citenamefont {Yang}, \citenamefont
  {Cao}, \citenamefont {Zhang}, \citenamefont {Ren}, \citenamefont {Cai},
  \citenamefont {Liu}, \citenamefont {Li}, \citenamefont {Shu}, \citenamefont
  {Huang}, \citenamefont {Deng}, \citenamefont {Li}, \citenamefont {Zhang},
  \citenamefont {Liu}, \citenamefont {Chen}, \citenamefont {Lu}, \citenamefont
  {Wang}, \citenamefont {Xu}, \citenamefont {Wang}, \citenamefont {Peng},
  \citenamefont {Ekert},\ and\ \citenamefont {Pan}}]{1120km_qkd}%
  \BibitemOpen
  \bibfield  {author} {\bibinfo {author} {\bibfnamefont {J.}~\bibnamefont
  {Yin}}, \bibinfo {author} {\bibfnamefont {Y.-H.}\ \bibnamefont {Li}},
  \bibinfo {author} {\bibfnamefont {S.-K.}\ \bibnamefont {Liao}}, \bibinfo
  {author} {\bibfnamefont {M.}~\bibnamefont {Yang}}, \bibinfo {author}
  {\bibfnamefont {Y.}~\bibnamefont {Cao}}, \bibinfo {author} {\bibfnamefont
  {L.}~\bibnamefont {Zhang}}, \bibinfo {author} {\bibfnamefont {J.-G.}\
  \bibnamefont {Ren}}, \bibinfo {author} {\bibfnamefont {W.-Q.}\ \bibnamefont
  {Cai}}, \bibinfo {author} {\bibfnamefont {W.-Y.}\ \bibnamefont {Liu}},
  \bibinfo {author} {\bibfnamefont {S.-L.}\ \bibnamefont {Li}}, \bibinfo
  {author} {\bibfnamefont {R.}~\bibnamefont {Shu}}, \bibinfo {author}
  {\bibfnamefont {Y.-M.}\ \bibnamefont {Huang}}, \bibinfo {author}
  {\bibfnamefont {L.}~\bibnamefont {Deng}}, \bibinfo {author} {\bibfnamefont
  {L.}~\bibnamefont {Li}}, \bibinfo {author} {\bibfnamefont {Q.}~\bibnamefont
  {Zhang}}, \bibinfo {author} {\bibfnamefont {N.-L.}\ \bibnamefont {Liu}},
  \bibinfo {author} {\bibfnamefont {Y.-A.}\ \bibnamefont {Chen}}, \bibinfo
  {author} {\bibfnamefont {C.-Y.}\ \bibnamefont {Lu}}, \bibinfo {author}
  {\bibfnamefont {X.-B.}\ \bibnamefont {Wang}}, \bibinfo {author}
  {\bibfnamefont {F.}~\bibnamefont {Xu}}, \bibinfo {author} {\bibfnamefont
  {J.-Y.}\ \bibnamefont {Wang}}, \bibinfo {author} {\bibfnamefont {C.-Z.}\
  \bibnamefont {Peng}}, \bibinfo {author} {\bibfnamefont {A.~K.}\ \bibnamefont
  {Ekert}},\ and\ \bibinfo {author} {\bibfnamefont {J.-W.}\ \bibnamefont
  {Pan}},\ }\bibfield  {title} {\bibinfo {title} {Entanglement-based secure
  quantum cryptography over 1,120 kilometres},\ }\href
  {https://doi.org/10.1038/s41586-020-2401-y} {\bibfield  {journal} {\bibinfo
  {journal} {Nature}\ }\textbf {\bibinfo {volume} {582}},\ \bibinfo {pages}
  {501} (\bibinfo {year} {2020})}\BibitemShut {NoStop}%
\bibitem [{\citenamefont {Chen}\ \emph {et~al.}(2021)\citenamefont {Chen},
  \citenamefont {Zhang}, \citenamefont {Chen}, \citenamefont {Cai},
  \citenamefont {Liao}, \citenamefont {Zhang}, \citenamefont {Chen},
  \citenamefont {Yin}, \citenamefont {Ren}, \citenamefont {Chen}, \citenamefont
  {Han}, \citenamefont {Yu}, \citenamefont {Liang}, \citenamefont {Zhou},
  \citenamefont {Yuan}, \citenamefont {Zhao}, \citenamefont {Wang},
  \citenamefont {Jiang}, \citenamefont {Zhang}, \citenamefont {Liu},
  \citenamefont {Li}, \citenamefont {Shen}, \citenamefont {Cao}, \citenamefont
  {Lu}, \citenamefont {Shu}, \citenamefont {Wang}, \citenamefont {Li},
  \citenamefont {Liu}, \citenamefont {Xu}, \citenamefont {Wang}, \citenamefont
  {Peng},\ and\ \citenamefont {Pan}}]{chen_integrated_2021}%
  \BibitemOpen
  \bibfield  {author} {\bibinfo {author} {\bibfnamefont {Y.-A.}\ \bibnamefont
  {Chen}}, \bibinfo {author} {\bibfnamefont {Q.}~\bibnamefont {Zhang}},
  \bibinfo {author} {\bibfnamefont {T.-Y.}\ \bibnamefont {Chen}}, \bibinfo
  {author} {\bibfnamefont {W.-Q.}\ \bibnamefont {Cai}}, \bibinfo {author}
  {\bibfnamefont {S.-K.}\ \bibnamefont {Liao}}, \bibinfo {author}
  {\bibfnamefont {J.}~\bibnamefont {Zhang}}, \bibinfo {author} {\bibfnamefont
  {K.}~\bibnamefont {Chen}}, \bibinfo {author} {\bibfnamefont {J.}~\bibnamefont
  {Yin}}, \bibinfo {author} {\bibfnamefont {J.-G.}\ \bibnamefont {Ren}},
  \bibinfo {author} {\bibfnamefont {Z.}~\bibnamefont {Chen}}, \bibinfo {author}
  {\bibfnamefont {S.-L.}\ \bibnamefont {Han}}, \bibinfo {author} {\bibfnamefont
  {Q.}~\bibnamefont {Yu}}, \bibinfo {author} {\bibfnamefont {K.}~\bibnamefont
  {Liang}}, \bibinfo {author} {\bibfnamefont {F.}~\bibnamefont {Zhou}},
  \bibinfo {author} {\bibfnamefont {X.}~\bibnamefont {Yuan}}, \bibinfo {author}
  {\bibfnamefont {M.-S.}\ \bibnamefont {Zhao}}, \bibinfo {author}
  {\bibfnamefont {T.-Y.}\ \bibnamefont {Wang}}, \bibinfo {author}
  {\bibfnamefont {X.}~\bibnamefont {Jiang}}, \bibinfo {author} {\bibfnamefont
  {L.}~\bibnamefont {Zhang}}, \bibinfo {author} {\bibfnamefont {W.-Y.}\
  \bibnamefont {Liu}}, \bibinfo {author} {\bibfnamefont {Y.}~\bibnamefont
  {Li}}, \bibinfo {author} {\bibfnamefont {Q.}~\bibnamefont {Shen}}, \bibinfo
  {author} {\bibfnamefont {Y.}~\bibnamefont {Cao}}, \bibinfo {author}
  {\bibfnamefont {C.-Y.}\ \bibnamefont {Lu}}, \bibinfo {author} {\bibfnamefont
  {R.}~\bibnamefont {Shu}}, \bibinfo {author} {\bibfnamefont {J.-Y.}\
  \bibnamefont {Wang}}, \bibinfo {author} {\bibfnamefont {L.}~\bibnamefont
  {Li}}, \bibinfo {author} {\bibfnamefont {N.-L.}\ \bibnamefont {Liu}},
  \bibinfo {author} {\bibfnamefont {F.}~\bibnamefont {Xu}}, \bibinfo {author}
  {\bibfnamefont {X.-B.}\ \bibnamefont {Wang}}, \bibinfo {author}
  {\bibfnamefont {C.-Z.}\ \bibnamefont {Peng}},\ and\ \bibinfo {author}
  {\bibfnamefont {J.-W.}\ \bibnamefont {Pan}},\ }\bibfield  {title} {\bibinfo
  {title} {An integrated space-to-ground quantum communication network over
  4,600 kilometres},\ }\href {https://doi.org/10.1038/s41586-020-03093-8}
  {\bibfield  {journal} {\bibinfo  {journal} {Nature}\ }\textbf {\bibinfo
  {volume} {589}},\ \bibinfo {pages} {214} (\bibinfo {year}
  {2021})}\BibitemShut {NoStop}%
\bibitem [{\citenamefont {Li}\ \emph {et~al.}(2023{\natexlab{a}})\citenamefont
  {Li}, \citenamefont {Zhang}, \citenamefont {Tan}, \citenamefont {Lu},
  \citenamefont {Liao}, \citenamefont {Huang}, \citenamefont {Li},
  \citenamefont {Wang}, \citenamefont {Mao}, \citenamefont {Yan}, \citenamefont
  {Li}, \citenamefont {Liu}, \citenamefont {Zhang}, \citenamefont {Peng},
  \citenamefont {You}, \citenamefont {Xu},\ and\ \citenamefont
  {Pan}}]{110M_qkd}%
  \BibitemOpen
  \bibfield  {author} {\bibinfo {author} {\bibfnamefont {W.}~\bibnamefont
  {Li}}, \bibinfo {author} {\bibfnamefont {L.}~\bibnamefont {Zhang}}, \bibinfo
  {author} {\bibfnamefont {H.}~\bibnamefont {Tan}}, \bibinfo {author}
  {\bibfnamefont {Y.}~\bibnamefont {Lu}}, \bibinfo {author} {\bibfnamefont
  {S.-K.}\ \bibnamefont {Liao}}, \bibinfo {author} {\bibfnamefont
  {J.}~\bibnamefont {Huang}}, \bibinfo {author} {\bibfnamefont
  {H.}~\bibnamefont {Li}}, \bibinfo {author} {\bibfnamefont {Z.}~\bibnamefont
  {Wang}}, \bibinfo {author} {\bibfnamefont {H.-K.}\ \bibnamefont {Mao}},
  \bibinfo {author} {\bibfnamefont {B.}~\bibnamefont {Yan}}, \bibinfo {author}
  {\bibfnamefont {Q.}~\bibnamefont {Li}}, \bibinfo {author} {\bibfnamefont
  {Y.}~\bibnamefont {Liu}}, \bibinfo {author} {\bibfnamefont {Q.}~\bibnamefont
  {Zhang}}, \bibinfo {author} {\bibfnamefont {C.-Z.}\ \bibnamefont {Peng}},
  \bibinfo {author} {\bibfnamefont {L.}~\bibnamefont {You}}, \bibinfo {author}
  {\bibfnamefont {F.}~\bibnamefont {Xu}},\ and\ \bibinfo {author}
  {\bibfnamefont {J.-W.}\ \bibnamefont {Pan}},\ }\bibfield  {title} {\bibinfo
  {title} {High-rate quantum key distribution exceeding 110 mb s–1},\ }\href
  {https://doi.org/10.1038/s41566-023-01166-4} {\bibfield  {journal} {\bibinfo
  {journal} {Nature Photonics}\ }\textbf {\bibinfo {volume} {17}},\ \bibinfo
  {pages} {416} (\bibinfo {year} {2023}{\natexlab{a}})}\BibitemShut {NoStop}%
\bibitem [{\citenamefont {Liu}\ \emph {et~al.}(2023)\citenamefont {Liu},
  \citenamefont {Zhang}, \citenamefont {Jiang}, \citenamefont {Chen},
  \citenamefont {Zhang}, \citenamefont {Pan}, \citenamefont {Ma}, \citenamefont
  {Dong}, \citenamefont {Xiong}, \citenamefont {Zhang}, \citenamefont {Li},
  \citenamefont {Wang}, \citenamefont {Wu}, \citenamefont {Chen}, \citenamefont
  {You}, \citenamefont {Wang}, \citenamefont {Zhang},\ and\ \citenamefont
  {Pan}}]{1000km_qkd}%
  \BibitemOpen
  \bibfield  {author} {\bibinfo {author} {\bibfnamefont {Y.}~\bibnamefont
  {Liu}}, \bibinfo {author} {\bibfnamefont {W.-J.}\ \bibnamefont {Zhang}},
  \bibinfo {author} {\bibfnamefont {C.}~\bibnamefont {Jiang}}, \bibinfo
  {author} {\bibfnamefont {J.-P.}\ \bibnamefont {Chen}}, \bibinfo {author}
  {\bibfnamefont {C.}~\bibnamefont {Zhang}}, \bibinfo {author} {\bibfnamefont
  {W.-X.}\ \bibnamefont {Pan}}, \bibinfo {author} {\bibfnamefont
  {D.}~\bibnamefont {Ma}}, \bibinfo {author} {\bibfnamefont {H.}~\bibnamefont
  {Dong}}, \bibinfo {author} {\bibfnamefont {J.-M.}\ \bibnamefont {Xiong}},
  \bibinfo {author} {\bibfnamefont {C.-J.}\ \bibnamefont {Zhang}}, \bibinfo
  {author} {\bibfnamefont {H.}~\bibnamefont {Li}}, \bibinfo {author}
  {\bibfnamefont {R.-C.}\ \bibnamefont {Wang}}, \bibinfo {author}
  {\bibfnamefont {J.}~\bibnamefont {Wu}}, \bibinfo {author} {\bibfnamefont
  {T.-Y.}\ \bibnamefont {Chen}}, \bibinfo {author} {\bibfnamefont
  {L.}~\bibnamefont {You}}, \bibinfo {author} {\bibfnamefont {X.-B.}\
  \bibnamefont {Wang}}, \bibinfo {author} {\bibfnamefont {Q.}~\bibnamefont
  {Zhang}},\ and\ \bibinfo {author} {\bibfnamefont {J.-W.}\ \bibnamefont
  {Pan}},\ }\bibfield  {title} {\bibinfo {title} {Experimental twin-field
  quantum key distribution over 1000 km fiber distance},\ }\href
  {https://doi.org/10.1103/PhysRevLett.130.210801} {\bibfield  {journal}
  {\bibinfo  {journal} {Phys. Rev. Lett.}\ }\textbf {\bibinfo {volume} {130}},\
  \bibinfo {pages} {210801} (\bibinfo {year} {2023})}\BibitemShut {NoStop}%
\bibitem [{\citenamefont {Wengerowsky}\ \emph {et~al.}(2018)\citenamefont
  {Wengerowsky}, \citenamefont {Joshi}, \citenamefont {Steinlechner},
  \citenamefont {Hübel},\ and\ \citenamefont {Ursin}}]{qkd_network_2}%
  \BibitemOpen
  \bibfield  {author} {\bibinfo {author} {\bibfnamefont {S.}~\bibnamefont
  {Wengerowsky}}, \bibinfo {author} {\bibfnamefont {S.~K.}\ \bibnamefont
  {Joshi}}, \bibinfo {author} {\bibfnamefont {F.}~\bibnamefont {Steinlechner}},
  \bibinfo {author} {\bibfnamefont {H.}~\bibnamefont {Hübel}},\ and\ \bibinfo
  {author} {\bibfnamefont {R.}~\bibnamefont {Ursin}},\ }\bibfield  {title}
  {\bibinfo {title} {An entanglement-based wavelength-multiplexed quantum
  communication network},\ }\href {https://doi.org/10.1038/s41586-018-0766-y}
  {\bibfield  {journal} {\bibinfo  {journal} {Nature}\ }\textbf {\bibinfo
  {volume} {564}},\ \bibinfo {pages} {225} (\bibinfo {year}
  {2018})}\BibitemShut {NoStop}%
\bibitem [{\citenamefont {Joshi}\ \emph {et~al.}(2020)\citenamefont {Joshi},
  \citenamefont {Aktas}, \citenamefont {Wengerowsky}, \citenamefont
  {Lončarić}, \citenamefont {Neumann}, \citenamefont {Liu}, \citenamefont
  {Scheidl}, \citenamefont {Lorenzo}, \citenamefont {Željko Samec},
  \citenamefont {Kling}, \citenamefont {Qiu}, \citenamefont {Razavi},
  \citenamefont {Stipčević}, \citenamefont {Rarity},\ and\ \citenamefont
  {Ursin}}]{qkd_network_3}%
  \BibitemOpen
  \bibfield  {author} {\bibinfo {author} {\bibfnamefont {S.~K.}\ \bibnamefont
  {Joshi}}, \bibinfo {author} {\bibfnamefont {D.}~\bibnamefont {Aktas}},
  \bibinfo {author} {\bibfnamefont {S.}~\bibnamefont {Wengerowsky}}, \bibinfo
  {author} {\bibfnamefont {M.}~\bibnamefont {Lončarić}}, \bibinfo {author}
  {\bibfnamefont {S.~P.}\ \bibnamefont {Neumann}}, \bibinfo {author}
  {\bibfnamefont {B.}~\bibnamefont {Liu}}, \bibinfo {author} {\bibfnamefont
  {T.}~\bibnamefont {Scheidl}}, \bibinfo {author} {\bibfnamefont {G.~C.}\
  \bibnamefont {Lorenzo}}, \bibinfo {author} {\bibnamefont {Željko Samec}},
  \bibinfo {author} {\bibfnamefont {L.}~\bibnamefont {Kling}}, \bibinfo
  {author} {\bibfnamefont {A.}~\bibnamefont {Qiu}}, \bibinfo {author}
  {\bibfnamefont {M.}~\bibnamefont {Razavi}}, \bibinfo {author} {\bibfnamefont
  {M.}~\bibnamefont {Stipčević}}, \bibinfo {author} {\bibfnamefont {J.~G.}\
  \bibnamefont {Rarity}},\ and\ \bibinfo {author} {\bibfnamefont
  {R.}~\bibnamefont {Ursin}},\ }\bibfield  {title} {\bibinfo {title} {A trusted
  node–free eight-user metropolitan quantum communication network},\ }\href
  {https://doi.org/10.1126/sciadv.aba0959} {\bibfield  {journal} {\bibinfo
  {journal} {Science Advances}\ }\textbf {\bibinfo {volume} {6}},\ \bibinfo
  {pages} {eaba0959} (\bibinfo {year} {2020})}\BibitemShut {NoStop}%
\bibitem [{\citenamefont {Wen}\ \emph {et~al.}(2022)\citenamefont {Wen},
  \citenamefont {Chen}, \citenamefont {Lu}, \citenamefont {Yan}, \citenamefont
  {Xue}, \citenamefont {Zhang}, \citenamefont {Lu}, \citenamefont {Zhu},\ and\
  \citenamefont {Ma}}]{qkd_network_1}%
  \BibitemOpen
  \bibfield  {author} {\bibinfo {author} {\bibfnamefont {W.}~\bibnamefont
  {Wen}}, \bibinfo {author} {\bibfnamefont {Z.}~\bibnamefont {Chen}}, \bibinfo
  {author} {\bibfnamefont {L.}~\bibnamefont {Lu}}, \bibinfo {author}
  {\bibfnamefont {W.}~\bibnamefont {Yan}}, \bibinfo {author} {\bibfnamefont
  {W.}~\bibnamefont {Xue}}, \bibinfo {author} {\bibfnamefont {P.}~\bibnamefont
  {Zhang}}, \bibinfo {author} {\bibfnamefont {Y.}~\bibnamefont {Lu}}, \bibinfo
  {author} {\bibfnamefont {S.}~\bibnamefont {Zhu}},\ and\ \bibinfo {author}
  {\bibfnamefont {X.-S.}\ \bibnamefont {Ma}},\ }\bibfield  {title} {\bibinfo
  {title} {Realizing an entanglement-based multiuser quantum network with
  integrated photonics},\ }\href
  {https://doi.org/10.1103/PhysRevApplied.18.024059} {\bibfield  {journal}
  {\bibinfo  {journal} {Phys. Rev. Appl.}\ }\textbf {\bibinfo {volume} {18}},\
  \bibinfo {pages} {024059} (\bibinfo {year} {2022})}\BibitemShut {NoStop}%
\bibitem [{\citenamefont {Yan}\ \emph {et~al.}(2025{\natexlab{a}})\citenamefont
  {Yan}, \citenamefont {Zheng}, \citenamefont {Wen}, \citenamefont {Lu},
  \citenamefont {Du}, \citenamefont {Lu}, \citenamefont {Zhu},\ and\
  \citenamefont {Ma}}]{yan_measurement-device-independent_2025}%
  \BibitemOpen
  \bibfield  {author} {\bibinfo {author} {\bibfnamefont {W.}~\bibnamefont
  {Yan}}, \bibinfo {author} {\bibfnamefont {X.}~\bibnamefont {Zheng}}, \bibinfo
  {author} {\bibfnamefont {W.}~\bibnamefont {Wen}}, \bibinfo {author}
  {\bibfnamefont {L.}~\bibnamefont {Lu}}, \bibinfo {author} {\bibfnamefont
  {Y.}~\bibnamefont {Du}}, \bibinfo {author} {\bibfnamefont {Y.-Q.}\
  \bibnamefont {Lu}}, \bibinfo {author} {\bibfnamefont {S.}~\bibnamefont
  {Zhu}},\ and\ \bibinfo {author} {\bibfnamefont {X.-S.}\ \bibnamefont {Ma}},\
  }\bibfield  {title} {\bibinfo {title} {A measurement-device-independent
  quantum key distribution network using optical frequency comb},\ }\href
  {https://doi.org/10.1038/s41534-025-01052-7} {\bibfield  {journal} {\bibinfo
  {journal} {npj Quantum Information}\ }\textbf {\bibinfo {volume} {11}},\
  \bibinfo {pages} {97} (\bibinfo {year} {2025}{\natexlab{a}})}\BibitemShut
  {NoStop}%
\bibitem [{\citenamefont {Bose}\ \emph {et~al.}(1998)\citenamefont {Bose},
  \citenamefont {Vedral},\ and\ \citenamefont {Knight}}]{qcc_concept_1}%
  \BibitemOpen
  \bibfield  {author} {\bibinfo {author} {\bibfnamefont {S.}~\bibnamefont
  {Bose}}, \bibinfo {author} {\bibfnamefont {V.}~\bibnamefont {Vedral}},\ and\
  \bibinfo {author} {\bibfnamefont {P.~L.}\ \bibnamefont {Knight}},\ }\bibfield
   {title} {\bibinfo {title} {Multiparticle generalization of entanglement
  swapping},\ }\href {https://doi.org/10.1103/PhysRevA.57.822} {\bibfield
  {journal} {\bibinfo  {journal} {Phys. Rev. A}\ }\textbf {\bibinfo {volume}
  {57}},\ \bibinfo {pages} {822} (\bibinfo {year} {1998})}\BibitemShut
  {NoStop}%
\bibitem [{\citenamefont {Chen}\ and\ \citenamefont
  {Lo}(2007)}]{qcc_concept_2}%
  \BibitemOpen
  \bibfield  {author} {\bibinfo {author} {\bibfnamefont {K.}~\bibnamefont
  {Chen}}\ and\ \bibinfo {author} {\bibfnamefont {H.-K.}\ \bibnamefont {Lo}},\
  }\bibfield  {title} {\bibinfo {title} {Multipartite quantum cryptographic
  protocols with noisy ghz states},\ }\href
  {https://dl.acm.org/doi/10.5555/2011742.2011743} {\bibfield  {journal}
  {\bibinfo  {journal} {Quantum Information and Computation}\ }\textbf
  {\bibinfo {volume} {7}} (\bibinfo {year} {2007})}\BibitemShut {NoStop}%
\bibitem [{\citenamefont {Murta}\ \emph {et~al.}(2020)\citenamefont {Murta},
  \citenamefont {Grasselli}, \citenamefont {Kampermann},\ and\ \citenamefont
  {Bruß}}]{qcc_review}%
  \BibitemOpen
  \bibfield  {author} {\bibinfo {author} {\bibfnamefont {G.}~\bibnamefont
  {Murta}}, \bibinfo {author} {\bibfnamefont {F.}~\bibnamefont {Grasselli}},
  \bibinfo {author} {\bibfnamefont {H.}~\bibnamefont {Kampermann}},\ and\
  \bibinfo {author} {\bibfnamefont {D.}~\bibnamefont {Bruß}},\ }\bibfield
  {title} {\bibinfo {title} {Quantum conference key agreement: A review},\
  }\href {https://doi.org/https://doi.org/10.1002/qute.202000025} {\bibfield
  {journal} {\bibinfo  {journal} {Advanced Quantum Technologies}\ }\textbf
  {\bibinfo {volume} {3}},\ \bibinfo {pages} {2000025} (\bibinfo {year}
  {2020})}\BibitemShut {NoStop}%
\bibitem [{\citenamefont {Augusiak}\ and\ \citenamefont
  {Horodecki}(2009)}]{qcc_gme_2}%
  \BibitemOpen
  \bibfield  {author} {\bibinfo {author} {\bibfnamefont {R.}~\bibnamefont
  {Augusiak}}\ and\ \bibinfo {author} {\bibfnamefont {P.}~\bibnamefont
  {Horodecki}},\ }\bibfield  {title} {\bibinfo {title} {Multipartite secret key
  distillation and bound entanglement},\ }\href
  {https://doi.org/10.1103/PhysRevA.80.042307} {\bibfield  {journal} {\bibinfo
  {journal} {Phys. Rev. A}\ }\textbf {\bibinfo {volume} {80}},\ \bibinfo
  {pages} {042307} (\bibinfo {year} {2009})}\BibitemShut {NoStop}%
\bibitem [{\citenamefont {Carrara}\ \emph {et~al.}(2021)\citenamefont
  {Carrara}, \citenamefont {Kampermann}, \citenamefont {Bru\ss{}},\ and\
  \citenamefont {Murta}}]{qcc_gme_3}%
  \BibitemOpen
  \bibfield  {author} {\bibinfo {author} {\bibfnamefont {G.}~\bibnamefont
  {Carrara}}, \bibinfo {author} {\bibfnamefont {H.}~\bibnamefont {Kampermann}},
  \bibinfo {author} {\bibfnamefont {D.}~\bibnamefont {Bru\ss{}}},\ and\
  \bibinfo {author} {\bibfnamefont {G.}~\bibnamefont {Murta}},\ }\bibfield
  {title} {\bibinfo {title} {Genuine multipartite entanglement is not a
  precondition for secure conference key agreement},\ }\href
  {https://doi.org/10.1103/PhysRevResearch.3.013264} {\bibfield  {journal}
  {\bibinfo  {journal} {Phys. Rev. Res.}\ }\textbf {\bibinfo {volume} {3}},\
  \bibinfo {pages} {013264} (\bibinfo {year} {2021})}\BibitemShut {NoStop}%
\bibitem [{\citenamefont {Das}\ \emph {et~al.}(2021)\citenamefont {Das},
  \citenamefont {B\"auml}, \citenamefont {Winczewski},\ and\ \citenamefont
  {Horodecki}}]{qcc_gme_1}%
  \BibitemOpen
  \bibfield  {author} {\bibinfo {author} {\bibfnamefont {S.}~\bibnamefont
  {Das}}, \bibinfo {author} {\bibfnamefont {S.}~\bibnamefont {B\"auml}},
  \bibinfo {author} {\bibfnamefont {M.}~\bibnamefont {Winczewski}},\ and\
  \bibinfo {author} {\bibfnamefont {K.}~\bibnamefont {Horodecki}},\ }\bibfield
  {title} {\bibinfo {title} {Universal limitations on quantum key distribution
  over a network},\ }\href {https://doi.org/10.1103/PhysRevX.11.041016}
  {\bibfield  {journal} {\bibinfo  {journal} {Phys. Rev. X}\ }\textbf {\bibinfo
  {volume} {11}},\ \bibinfo {pages} {041016} (\bibinfo {year}
  {2021})}\BibitemShut {NoStop}%
\bibitem [{\citenamefont {Epping}\ \emph {et~al.}(2017)\citenamefont {Epping},
  \citenamefont {Kampermann}, \citenamefont {macchiavello},\ and\ \citenamefont
  {Bruß}}]{qcc_eb_theory_2}%
  \BibitemOpen
  \bibfield  {author} {\bibinfo {author} {\bibfnamefont {M.}~\bibnamefont
  {Epping}}, \bibinfo {author} {\bibfnamefont {H.}~\bibnamefont {Kampermann}},
  \bibinfo {author} {\bibfnamefont {C.}~\bibnamefont {macchiavello}},\ and\
  \bibinfo {author} {\bibfnamefont {D.}~\bibnamefont {Bruß}},\ }\bibfield
  {title} {\bibinfo {title} {Multi-partite entanglement can speed up quantum
  key distribution in networks},\ }\href
  {https://doi.org/10.1088/1367-2630/aa8487} {\bibfield  {journal} {\bibinfo
  {journal} {New Journal of Physics}\ }\textbf {\bibinfo {volume} {19}},\
  \bibinfo {pages} {093012} (\bibinfo {year} {2017})}\BibitemShut {NoStop}%
\bibitem [{\citenamefont {Grasselli}\ \emph {et~al.}(2018)\citenamefont
  {Grasselli}, \citenamefont {Kampermann},\ and\ \citenamefont
  {Bruß}}]{qcc_eb_theory_1}%
  \BibitemOpen
  \bibfield  {author} {\bibinfo {author} {\bibfnamefont {F.}~\bibnamefont
  {Grasselli}}, \bibinfo {author} {\bibfnamefont {H.}~\bibnamefont
  {Kampermann}},\ and\ \bibinfo {author} {\bibfnamefont {D.}~\bibnamefont
  {Bruß}},\ }\bibfield  {title} {\bibinfo {title} {Finite-key effects in
  multipartite quantum key distribution protocols},\ }\href
  {https://doi.org/10.1088/1367-2630/aaec34} {\bibfield  {journal} {\bibinfo
  {journal} {New Journal of Physics}\ }\textbf {\bibinfo {volume} {20}},\
  \bibinfo {pages} {113014} (\bibinfo {year} {2018})}\BibitemShut {NoStop}%
\bibitem [{\citenamefont {Fu}\ \emph {et~al.}(2015)\citenamefont {Fu},
  \citenamefont {Yin}, \citenamefont {Chen},\ and\ \citenamefont
  {Chen}}]{pol_qcc}%
  \BibitemOpen
  \bibfield  {author} {\bibinfo {author} {\bibfnamefont {Y.}~\bibnamefont
  {Fu}}, \bibinfo {author} {\bibfnamefont {H.-L.}\ \bibnamefont {Yin}},
  \bibinfo {author} {\bibfnamefont {T.-Y.}\ \bibnamefont {Chen}},\ and\
  \bibinfo {author} {\bibfnamefont {Z.-B.}\ \bibnamefont {Chen}},\ }\bibfield
  {title} {\bibinfo {title} {Long-distance measurement-device-independent
  multiparty quantum communication},\ }\href
  {https://doi.org/10.1103/PhysRevLett.114.090501} {\bibfield  {journal}
  {\bibinfo  {journal} {Phys. Rev. Lett.}\ }\textbf {\bibinfo {volume} {114}},\
  \bibinfo {pages} {090501} (\bibinfo {year} {2015})}\BibitemShut {NoStop}%
\bibitem [{\citenamefont {Grasselli}\ \emph {et~al.}(2019)\citenamefont
  {Grasselli}, \citenamefont {Kampermann},\ and\ \citenamefont
  {Bruß}}]{w_qcc_1}%
  \BibitemOpen
  \bibfield  {author} {\bibinfo {author} {\bibfnamefont {F.}~\bibnamefont
  {Grasselli}}, \bibinfo {author} {\bibfnamefont {H.}~\bibnamefont
  {Kampermann}},\ and\ \bibinfo {author} {\bibfnamefont {D.}~\bibnamefont
  {Bruß}},\ }\bibfield  {title} {\bibinfo {title} {Conference key agreement
  with single-photon interference},\ }\href
  {https://doi.org/10.1088/1367-2630/ab573e} {\bibfield  {journal} {\bibinfo
  {journal} {New Journal of Physics}\ }\textbf {\bibinfo {volume} {21}},\
  \bibinfo {pages} {123002} (\bibinfo {year} {2019})}\BibitemShut {NoStop}%
\bibitem [{\citenamefont {Zhao}\ \emph {et~al.}(2020)\citenamefont {Zhao},
  \citenamefont {Zeng}, \citenamefont {Cao}, \citenamefont {Xu}, \citenamefont
  {Zhen}, \citenamefont {Ma}, \citenamefont {Li}, \citenamefont {Liu},\ and\
  \citenamefont {Chen}}]{pm_qcc}%
  \BibitemOpen
  \bibfield  {author} {\bibinfo {author} {\bibfnamefont {S.}~\bibnamefont
  {Zhao}}, \bibinfo {author} {\bibfnamefont {P.}~\bibnamefont {Zeng}}, \bibinfo
  {author} {\bibfnamefont {W.-F.}\ \bibnamefont {Cao}}, \bibinfo {author}
  {\bibfnamefont {X.-Y.}\ \bibnamefont {Xu}}, \bibinfo {author} {\bibfnamefont
  {Y.-Z.}\ \bibnamefont {Zhen}}, \bibinfo {author} {\bibfnamefont
  {X.}~\bibnamefont {Ma}}, \bibinfo {author} {\bibfnamefont {L.}~\bibnamefont
  {Li}}, \bibinfo {author} {\bibfnamefont {N.-L.}\ \bibnamefont {Liu}},\ and\
  \bibinfo {author} {\bibfnamefont {K.}~\bibnamefont {Chen}},\ }\bibfield
  {title} {\bibinfo {title} {Phase-matching quantum cryptographic
  conferencing},\ }\href {https://doi.org/10.1103/PhysRevApplied.14.024010}
  {\bibfield  {journal} {\bibinfo  {journal} {Phys. Rev. Appl.}\ }\textbf
  {\bibinfo {volume} {14}},\ \bibinfo {pages} {024010} (\bibinfo {year}
  {2020})}\BibitemShut {NoStop}%
\bibitem [{\citenamefont {Cao}\ \emph {et~al.}(2021{\natexlab{a}})\citenamefont
  {Cao}, \citenamefont {Gu}, \citenamefont {Lu}, \citenamefont {Yin},\ and\
  \citenamefont {Chen}}]{cow_qcc_2}%
  \BibitemOpen
  \bibfield  {author} {\bibinfo {author} {\bibfnamefont {X.-Y.}\ \bibnamefont
  {Cao}}, \bibinfo {author} {\bibfnamefont {J.}~\bibnamefont {Gu}}, \bibinfo
  {author} {\bibfnamefont {Y.-S.}\ \bibnamefont {Lu}}, \bibinfo {author}
  {\bibfnamefont {H.-L.}\ \bibnamefont {Yin}},\ and\ \bibinfo {author}
  {\bibfnamefont {Z.-B.}\ \bibnamefont {Chen}},\ }\bibfield  {title} {\bibinfo
  {title} {Coherent one-way quantum conference key agreement based on twin
  field},\ }\href {https://doi.org/10.1088/1367-2630/abef98} {\bibfield
  {journal} {\bibinfo  {journal} {New Journal of Physics}\ }\textbf {\bibinfo
  {volume} {23}},\ \bibinfo {pages} {043002} (\bibinfo {year}
  {2021}{\natexlab{a}})}\BibitemShut {NoStop}%
\bibitem [{\citenamefont {Cao}\ \emph {et~al.}(2021{\natexlab{b}})\citenamefont
  {Cao}, \citenamefont {Lu}, \citenamefont {Li}, \citenamefont {Gu},
  \citenamefont {Yin},\ and\ \citenamefont {Chen}}]{cow_qcc_1}%
  \BibitemOpen
  \bibfield  {author} {\bibinfo {author} {\bibfnamefont {X.-Y.}\ \bibnamefont
  {Cao}}, \bibinfo {author} {\bibfnamefont {Y.-S.}\ \bibnamefont {Lu}},
  \bibinfo {author} {\bibfnamefont {Z.}~\bibnamefont {Li}}, \bibinfo {author}
  {\bibfnamefont {J.}~\bibnamefont {Gu}}, \bibinfo {author} {\bibfnamefont
  {H.-L.}\ \bibnamefont {Yin}},\ and\ \bibinfo {author} {\bibfnamefont {Z.-B.}\
  \bibnamefont {Chen}},\ }\bibfield  {title} {\bibinfo {title} {High key rate
  quantum conference key agreement with unconditional security},\ }\href
  {https://doi.org/10.1109/ACCESS.2021.3113939} {\bibfield  {journal} {\bibinfo
   {journal} {IEEE Access}\ }\textbf {\bibinfo {volume} {9}},\ \bibinfo {pages}
  {128870} (\bibinfo {year} {2021}{\natexlab{b}})}\BibitemShut {NoStop}%
\bibitem [{\citenamefont {Bai}\ \emph {et~al.}(2022)\citenamefont {Bai},
  \citenamefont {Xie}, \citenamefont {Li}, \citenamefont {Yin},\ and\
  \citenamefont {Chen}}]{cow_qcc_3}%
  \BibitemOpen
  \bibfield  {author} {\bibinfo {author} {\bibfnamefont {J.-L.}\ \bibnamefont
  {Bai}}, \bibinfo {author} {\bibfnamefont {Y.-M.}\ \bibnamefont {Xie}},
  \bibinfo {author} {\bibfnamefont {Z.}~\bibnamefont {Li}}, \bibinfo {author}
  {\bibfnamefont {H.-L.}\ \bibnamefont {Yin}},\ and\ \bibinfo {author}
  {\bibfnamefont {Z.-B.}\ \bibnamefont {Chen}},\ }\bibfield  {title} {\bibinfo
  {title} {Post-matching quantum conference key agreement},\ }\href
  {https://doi.org/10.1364/OE.460725} {\bibfield  {journal} {\bibinfo
  {journal} {Optics Express}\ }\textbf {\bibinfo {volume} {30}},\ \bibinfo
  {pages} {28865} (\bibinfo {year} {2022})}\BibitemShut {NoStop}%
\bibitem [{\citenamefont {Carrara}\ \emph {et~al.}(2023)\citenamefont
  {Carrara}, \citenamefont {Murta},\ and\ \citenamefont {Grasselli}}]{w_qcc_2}%
  \BibitemOpen
  \bibfield  {author} {\bibinfo {author} {\bibfnamefont {G.}~\bibnamefont
  {Carrara}}, \bibinfo {author} {\bibfnamefont {G.}~\bibnamefont {Murta}},\
  and\ \bibinfo {author} {\bibfnamefont {F.}~\bibnamefont {Grasselli}},\
  }\bibfield  {title} {\bibinfo {title} {Overcoming fundamental bounds on
  quantum conference key agreement},\ }\href
  {https://doi.org/10.1103/PhysRevApplied.19.064017} {\bibfield  {journal}
  {\bibinfo  {journal} {Phys. Rev. Appl.}\ }\textbf {\bibinfo {volume} {19}},\
  \bibinfo {pages} {064017} (\bibinfo {year} {2023})}\BibitemShut {NoStop}%
\bibitem [{\citenamefont {Lu}\ \emph {et~al.}(2025{\natexlab{a}})\citenamefont
  {Lu}, \citenamefont {Yin}, \citenamefont {Xie}, \citenamefont {Fu},\ and\
  \citenamefont {Chen}}]{lu2024repeaterlike}%
  \BibitemOpen
  \bibfield  {author} {\bibinfo {author} {\bibfnamefont {Y.-S.}\ \bibnamefont
  {Lu}}, \bibinfo {author} {\bibfnamefont {H.-L.}\ \bibnamefont {Yin}},
  \bibinfo {author} {\bibfnamefont {Y.-M.}\ \bibnamefont {Xie}}, \bibinfo
  {author} {\bibfnamefont {Y.}~\bibnamefont {Fu}},\ and\ \bibinfo {author}
  {\bibfnamefont {Z.-B.}\ \bibnamefont {Chen}},\ }\bibfield  {title} {\bibinfo
  {title} {Repeater-like asynchronous measurement-device-independent quantum
  conference key agreement},\ }\href {https://doi.org/10.1088/1361-6633/addeec}
  {\bibfield  {journal} {\bibinfo  {journal} {Reports on Progress in Physics}\
  }\textbf {\bibinfo {volume} {88}},\ \bibinfo {pages} {067901} (\bibinfo
  {year} {2025}{\natexlab{a}})}\BibitemShut {NoStop}%
\bibitem [{\citenamefont {Xie}\ \emph {et~al.}(2024)\citenamefont {Xie},
  \citenamefont {Lu}, \citenamefont {Fu}, \citenamefont {Yin},\ and\
  \citenamefont {Chen}}]{xie2024multifield}%
  \BibitemOpen
  \bibfield  {author} {\bibinfo {author} {\bibfnamefont {Y.-M.}\ \bibnamefont
  {Xie}}, \bibinfo {author} {\bibfnamefont {Y.-S.}\ \bibnamefont {Lu}},
  \bibinfo {author} {\bibfnamefont {Y.}~\bibnamefont {Fu}}, \bibinfo {author}
  {\bibfnamefont {H.-L.}\ \bibnamefont {Yin}},\ and\ \bibinfo {author}
  {\bibfnamefont {Z.-B.}\ \bibnamefont {Chen}},\ }\bibfield  {title} {\bibinfo
  {title} {Multi-field quantum conferencing overcomes the network capacity
  limit},\ }\href {https://doi.org/10.1038/s42005-024-01894-1} {\bibfield
  {journal} {\bibinfo  {journal} {Communications Physics}\ }\textbf {\bibinfo
  {volume} {7}},\ \bibinfo {pages} {410} (\bibinfo {year} {2024})}\BibitemShut
  {NoStop}%
\bibitem [{\citenamefont {Proietti}\ \emph {et~al.}(2021)\citenamefont
  {Proietti}, \citenamefont {Ho}, \citenamefont {Grasselli}, \citenamefont
  {Barrow}, \citenamefont {Malik},\ and\ \citenamefont
  {Fedrizzi}}]{qcc_eb_exp_1}%
  \BibitemOpen
  \bibfield  {author} {\bibinfo {author} {\bibfnamefont {M.}~\bibnamefont
  {Proietti}}, \bibinfo {author} {\bibfnamefont {J.}~\bibnamefont {Ho}},
  \bibinfo {author} {\bibfnamefont {F.}~\bibnamefont {Grasselli}}, \bibinfo
  {author} {\bibfnamefont {P.}~\bibnamefont {Barrow}}, \bibinfo {author}
  {\bibfnamefont {M.}~\bibnamefont {Malik}},\ and\ \bibinfo {author}
  {\bibfnamefont {A.}~\bibnamefont {Fedrizzi}},\ }\bibfield  {title} {\bibinfo
  {title} {Experimental quantum conference key agreement},\ }\href
  {https://doi.org/10.1126/sciadv.abe0395} {\bibfield  {journal} {\bibinfo
  {journal} {Science Advances}\ }\textbf {\bibinfo {volume} {7}},\ \bibinfo
  {pages} {eabe0395} (\bibinfo {year} {2021})}\BibitemShut {NoStop}%
\bibitem [{\citenamefont {Pickston}\ \emph {et~al.}(2023)\citenamefont
  {Pickston}, \citenamefont {Ho}, \citenamefont {Ulibarrena}, \citenamefont
  {Grasselli}, \citenamefont {Proietti}, \citenamefont {Morrison},
  \citenamefont {Barrow}, \citenamefont {Graffitti},\ and\ \citenamefont
  {Fedrizzi}}]{qcc_eb_exp_2}%
  \BibitemOpen
  \bibfield  {author} {\bibinfo {author} {\bibfnamefont {A.}~\bibnamefont
  {Pickston}}, \bibinfo {author} {\bibfnamefont {J.}~\bibnamefont {Ho}},
  \bibinfo {author} {\bibfnamefont {A.}~\bibnamefont {Ulibarrena}}, \bibinfo
  {author} {\bibfnamefont {F.}~\bibnamefont {Grasselli}}, \bibinfo {author}
  {\bibfnamefont {M.}~\bibnamefont {Proietti}}, \bibinfo {author}
  {\bibfnamefont {C.~L.}\ \bibnamefont {Morrison}}, \bibinfo {author}
  {\bibfnamefont {P.}~\bibnamefont {Barrow}}, \bibinfo {author} {\bibfnamefont
  {F.}~\bibnamefont {Graffitti}},\ and\ \bibinfo {author} {\bibfnamefont
  {A.}~\bibnamefont {Fedrizzi}},\ }\bibfield  {title} {\bibinfo {title}
  {Conference key agreement in a quantum network},\ }\href
  {https://doi.org/10.1038/s41534-023-00750-4} {\bibfield  {journal} {\bibinfo
  {journal} {npj Quantum Information}\ }\textbf {\bibinfo {volume} {9}},\
  \bibinfo {pages} {82} (\bibinfo {year} {2023})}\BibitemShut {NoStop}%
\bibitem [{\citenamefont {Pirandola}\ \emph {et~al.}(2017)\citenamefont
  {Pirandola}, \citenamefont {Laurenza}, \citenamefont {Ottaviani},\ and\
  \citenamefont {Banchi}}]{pirandola_fundamental_2017}%
  \BibitemOpen
  \bibfield  {author} {\bibinfo {author} {\bibfnamefont {S.}~\bibnamefont
  {Pirandola}}, \bibinfo {author} {\bibfnamefont {R.}~\bibnamefont {Laurenza}},
  \bibinfo {author} {\bibfnamefont {C.}~\bibnamefont {Ottaviani}},\ and\
  \bibinfo {author} {\bibfnamefont {L.}~\bibnamefont {Banchi}},\ }\bibfield
  {title} {\bibinfo {title} {Fundamental limits of repeaterless quantum
  communications},\ }\href {https://doi.org/10.1038/ncomms15043} {\bibfield
  {journal} {\bibinfo  {journal} {Nature Communications}\ }\textbf {\bibinfo
  {volume} {8}},\ \bibinfo {pages} {15043} (\bibinfo {year}
  {2017})}\BibitemShut {NoStop}%
\bibitem [{\citenamefont {Yang}\ \emph {et~al.}(2024)\citenamefont {Yang},
  \citenamefont {Mao}, \citenamefont {Chen}, \citenamefont {Dong},
  \citenamefont {Zhu}, \citenamefont {Wu},\ and\ \citenamefont
  {Li}}]{PhysRevLett.133.210803}%
  \BibitemOpen
  \bibfield  {author} {\bibinfo {author} {\bibfnamefont {K.-X.}\ \bibnamefont
  {Yang}}, \bibinfo {author} {\bibfnamefont {Y.-L.}\ \bibnamefont {Mao}},
  \bibinfo {author} {\bibfnamefont {H.}~\bibnamefont {Chen}}, \bibinfo {author}
  {\bibfnamefont {X.}~\bibnamefont {Dong}}, \bibinfo {author} {\bibfnamefont
  {J.}~\bibnamefont {Zhu}}, \bibinfo {author} {\bibfnamefont {J.}~\bibnamefont
  {Wu}},\ and\ \bibinfo {author} {\bibfnamefont {Z.-D.}\ \bibnamefont {Li}},\
  }\bibfield  {title} {\bibinfo {title} {Experimental
  measurement-device-independent quantum conference key agreement},\ }\href
  {https://doi.org/10.1103/PhysRevLett.133.210803} {\bibfield  {journal}
  {\bibinfo  {journal} {Phys. Rev. Lett.}\ }\textbf {\bibinfo {volume} {133}},\
  \bibinfo {pages} {210803} (\bibinfo {year} {2024})}\BibitemShut {NoStop}%
\bibitem [{\citenamefont {Du}\ \emph {et~al.}(2025)\citenamefont {Du},
  \citenamefont {Liu}, \citenamefont {Yang}, \citenamefont {Zheng},
  \citenamefont {Zhu},\ and\ \citenamefont {Ma}}]{du_experimental_2025}%
  \BibitemOpen
  \bibfield  {author} {\bibinfo {author} {\bibfnamefont {Y.}~\bibnamefont
  {Du}}, \bibinfo {author} {\bibfnamefont {Y.}~\bibnamefont {Liu}}, \bibinfo
  {author} {\bibfnamefont {C.}~\bibnamefont {Yang}}, \bibinfo {author}
  {\bibfnamefont {X.}~\bibnamefont {Zheng}}, \bibinfo {author} {\bibfnamefont
  {S.}~\bibnamefont {Zhu}},\ and\ \bibinfo {author} {\bibfnamefont {X.-S.}\
  \bibnamefont {Ma}},\ }\bibfield  {title} {\bibinfo {title} {Experimental
  {Measurement}-{Device}-{Independent} {Quantum} {Cryptographic}
  {Conferencing}},\ }\href {https://doi.org/10.1103/PhysRevLett.134.040802}
  {\bibfield  {journal} {\bibinfo  {journal} {Phys. Rev. Lett.}\ }\textbf
  {\bibinfo {volume} {134}},\ \bibinfo {pages} {040802} (\bibinfo {year}
  {2025})}\BibitemShut {NoStop}%
\bibitem [{\citenamefont {Zou}\ \emph {et~al.}(2026)\citenamefont {Zou},
  \citenamefont {Li}, \citenamefont {Zhao}, \citenamefont {Mao}, \citenamefont
  {Qin}, \citenamefont {Jiang}, \citenamefont {Chen},\ and\ \citenamefont
  {Pan}}]{pm-qcc-exp}%
  \BibitemOpen
  \bibfield  {author} {\bibinfo {author} {\bibfnamefont {M.}~\bibnamefont
  {Zou}}, \bibinfo {author} {\bibfnamefont {B.-C.}\ \bibnamefont {Li}},
  \bibinfo {author} {\bibfnamefont {S.}~\bibnamefont {Zhao}}, \bibinfo {author}
  {\bibfnamefont {Y.}~\bibnamefont {Mao}}, \bibinfo {author} {\bibfnamefont
  {D.}~\bibnamefont {Qin}}, \bibinfo {author} {\bibfnamefont {X.}~\bibnamefont
  {Jiang}}, \bibinfo {author} {\bibfnamefont {T.-Y.}\ \bibnamefont {Chen}},\
  and\ \bibinfo {author} {\bibfnamefont {J.-W.}\ \bibnamefont {Pan}},\
  }\bibfield  {title} {\bibinfo {title} {Experimental phase-matching quantum
  cryptographic conferencing in symmetric and asymmetric fiber channels},\
  }\href {https://doi.org/10.1103/mlgs-g354} {\bibfield  {journal} {\bibinfo
  {journal} {Phys. Rev. Lett.}\ }\textbf {\bibinfo {volume} {136}},\ \bibinfo
  {pages} {020801} (\bibinfo {year} {2026})}\BibitemShut {NoStop}%
\bibitem [{\citenamefont {Zhu}\ \emph {et~al.}(2023)\citenamefont {Zhu},
  \citenamefont {Huang}, \citenamefont {Liu}, \citenamefont {Zeng},
  \citenamefont {Zou}, \citenamefont {Dai}, \citenamefont {Tang}, \citenamefont
  {Li}, \citenamefont {You}, \citenamefont {Wang}, \citenamefont {Chen},
  \citenamefont {Ma}, \citenamefont {Chen},\ and\ \citenamefont
  {Pan}}]{mp_zhu_experimental_2023}%
  \BibitemOpen
  \bibfield  {author} {\bibinfo {author} {\bibfnamefont {H.-T.}\ \bibnamefont
  {Zhu}}, \bibinfo {author} {\bibfnamefont {Y.}~\bibnamefont {Huang}}, \bibinfo
  {author} {\bibfnamefont {H.}~\bibnamefont {Liu}}, \bibinfo {author}
  {\bibfnamefont {P.}~\bibnamefont {Zeng}}, \bibinfo {author} {\bibfnamefont
  {M.}~\bibnamefont {Zou}}, \bibinfo {author} {\bibfnamefont {Y.}~\bibnamefont
  {Dai}}, \bibinfo {author} {\bibfnamefont {S.}~\bibnamefont {Tang}}, \bibinfo
  {author} {\bibfnamefont {H.}~\bibnamefont {Li}}, \bibinfo {author}
  {\bibfnamefont {L.}~\bibnamefont {You}}, \bibinfo {author} {\bibfnamefont
  {Z.}~\bibnamefont {Wang}}, \bibinfo {author} {\bibfnamefont {Y.-A.}\
  \bibnamefont {Chen}}, \bibinfo {author} {\bibfnamefont {X.}~\bibnamefont
  {Ma}}, \bibinfo {author} {\bibfnamefont {T.-Y.}\ \bibnamefont {Chen}},\ and\
  \bibinfo {author} {\bibfnamefont {J.-W.}\ \bibnamefont {Pan}},\ }\bibfield
  {title} {\bibinfo {title} {Experimental {Mode}-{Pairing}
  {Measurement}-{Device}-{Independent} {Quantum} {Key} {Distribution} without
  {Global} {Phase} {Locking}},\ }\href
  {https://doi.org/10.1103/PhysRevLett.130.030801} {\bibfield  {journal}
  {\bibinfo  {journal} {Phys. Rev. Lett.}\ }\textbf {\bibinfo {volume} {130}},\
  \bibinfo {pages} {030801} (\bibinfo {year} {2023})}\BibitemShut {NoStop}%
\bibitem [{\citenamefont {Zhou}\ \emph {et~al.}(2023)\citenamefont {Zhou},
  \citenamefont {Lin}, \citenamefont {Xie}, \citenamefont {Lu}, \citenamefont
  {Jing}, \citenamefont {Yin},\ and\ \citenamefont
  {Yuan}}]{mp_zhou_experimental_2023}%
  \BibitemOpen
  \bibfield  {author} {\bibinfo {author} {\bibfnamefont {L.}~\bibnamefont
  {Zhou}}, \bibinfo {author} {\bibfnamefont {J.}~\bibnamefont {Lin}}, \bibinfo
  {author} {\bibfnamefont {Y.-M.}\ \bibnamefont {Xie}}, \bibinfo {author}
  {\bibfnamefont {Y.-S.}\ \bibnamefont {Lu}}, \bibinfo {author} {\bibfnamefont
  {Y.}~\bibnamefont {Jing}}, \bibinfo {author} {\bibfnamefont {H.-L.}\
  \bibnamefont {Yin}},\ and\ \bibinfo {author} {\bibfnamefont {Z.}~\bibnamefont
  {Yuan}},\ }\bibfield  {title} {\bibinfo {title} {Experimental {Quantum}
  {Communication} {Overcomes} the {Rate}-{Loss} {Limit} without {Global}
  {Phase} {Tracking}},\ }\href {https://doi.org/10.1103/PhysRevLett.130.250801}
  {\bibfield  {journal} {\bibinfo  {journal} {Phys. Rev. Lett.}\ }\textbf
  {\bibinfo {volume} {130}},\ \bibinfo {pages} {250801} (\bibinfo {year}
  {2023})}\BibitemShut {NoStop}%
\bibitem [{\citenamefont {Zhu}\ \emph {et~al.}(2024)\citenamefont {Zhu},
  \citenamefont {Huang}, \citenamefont {Pan}, \citenamefont {Zhou},
  \citenamefont {Tang}, \citenamefont {He}, \citenamefont {Cheng},
  \citenamefont {Jin}, \citenamefont {Zou}, \citenamefont {Tang}, \citenamefont
  {Ma}, \citenamefont {Chen},\ and\ \citenamefont {Pan}}]{mp_zhu_field_2024}%
  \BibitemOpen
  \bibfield  {author} {\bibinfo {author} {\bibfnamefont {H.-T.}\ \bibnamefont
  {Zhu}}, \bibinfo {author} {\bibfnamefont {Y.}~\bibnamefont {Huang}}, \bibinfo
  {author} {\bibfnamefont {W.-X.}\ \bibnamefont {Pan}}, \bibinfo {author}
  {\bibfnamefont {C.-W.}\ \bibnamefont {Zhou}}, \bibinfo {author}
  {\bibfnamefont {J.}~\bibnamefont {Tang}}, \bibinfo {author} {\bibfnamefont
  {H.}~\bibnamefont {He}}, \bibinfo {author} {\bibfnamefont {M.}~\bibnamefont
  {Cheng}}, \bibinfo {author} {\bibfnamefont {X.}~\bibnamefont {Jin}}, \bibinfo
  {author} {\bibfnamefont {M.}~\bibnamefont {Zou}}, \bibinfo {author}
  {\bibfnamefont {S.}~\bibnamefont {Tang}}, \bibinfo {author} {\bibfnamefont
  {X.}~\bibnamefont {Ma}}, \bibinfo {author} {\bibfnamefont {T.-Y.}\
  \bibnamefont {Chen}},\ and\ \bibinfo {author} {\bibfnamefont {J.-W.}\
  \bibnamefont {Pan}},\ }\bibfield  {title} {\bibinfo {title} {Field test of
  mode-pairing quantum key distribution},\ }\href
  {https://doi.org/10.1364/OPTICA.520697} {\bibfield  {journal} {\bibinfo
  {journal} {Optica}\ }\textbf {\bibinfo {volume} {11}},\ \bibinfo {pages}
  {883} (\bibinfo {year} {2024})}\BibitemShut {NoStop}%
\bibitem [{\citenamefont {Zhang}\ \emph {et~al.}(2025)\citenamefont {Zhang},
  \citenamefont {Li}, \citenamefont {Pan}, \citenamefont {Lu}, \citenamefont
  {Li}, \citenamefont {Li}, \citenamefont {Huang}, \citenamefont {Ma},
  \citenamefont {Xu},\ and\ \citenamefont {Pan}}]{mp_zhang_experimental_2025}%
  \BibitemOpen
  \bibfield  {author} {\bibinfo {author} {\bibfnamefont {L.}~\bibnamefont
  {Zhang}}, \bibinfo {author} {\bibfnamefont {W.}~\bibnamefont {Li}}, \bibinfo
  {author} {\bibfnamefont {J.}~\bibnamefont {Pan}}, \bibinfo {author}
  {\bibfnamefont {Y.}~\bibnamefont {Lu}}, \bibinfo {author} {\bibfnamefont
  {W.}~\bibnamefont {Li}}, \bibinfo {author} {\bibfnamefont {Z.-P.}\
  \bibnamefont {Li}}, \bibinfo {author} {\bibfnamefont {Y.}~\bibnamefont
  {Huang}}, \bibinfo {author} {\bibfnamefont {X.}~\bibnamefont {Ma}}, \bibinfo
  {author} {\bibfnamefont {F.}~\bibnamefont {Xu}},\ and\ \bibinfo {author}
  {\bibfnamefont {J.-W.}\ \bibnamefont {Pan}},\ }\bibfield  {title} {\bibinfo
  {title} {Experimental {Mode}-{Pairing} {Quantum} {Key} {Distribution}
  {Surpassing} the {Repeaterless} {Bound}},\ }\href
  {https://doi.org/10.1103/PhysRevX.15.021037} {\bibfield  {journal} {\bibinfo
  {journal} {Physical Review X}\ }\textbf {\bibinfo {volume} {15}},\ \bibinfo
  {pages} {021037} (\bibinfo {year} {2025})}\BibitemShut {NoStop}%
\bibitem [{\citenamefont {Shao}\ \emph {et~al.}(2025)\citenamefont {Shao},
  \citenamefont {Zhou}, \citenamefont {Lin}, \citenamefont {Minder},
  \citenamefont {Ge}, \citenamefont {Xie}, \citenamefont {Shen}, \citenamefont
  {Yan}, \citenamefont {Yin},\ and\ \citenamefont
  {Yuan}}]{mp_shao_high-rate_2025}%
  \BibitemOpen
  \bibfield  {author} {\bibinfo {author} {\bibfnamefont {S.-F.}\ \bibnamefont
  {Shao}}, \bibinfo {author} {\bibfnamefont {L.}~\bibnamefont {Zhou}}, \bibinfo
  {author} {\bibfnamefont {J.}~\bibnamefont {Lin}}, \bibinfo {author}
  {\bibfnamefont {M.}~\bibnamefont {Minder}}, \bibinfo {author} {\bibfnamefont
  {C.}~\bibnamefont {Ge}}, \bibinfo {author} {\bibfnamefont {Y.-M.}\
  \bibnamefont {Xie}}, \bibinfo {author} {\bibfnamefont {A.}~\bibnamefont
  {Shen}}, \bibinfo {author} {\bibfnamefont {Z.}~\bibnamefont {Yan}}, \bibinfo
  {author} {\bibfnamefont {H.-L.}\ \bibnamefont {Yin}},\ and\ \bibinfo {author}
  {\bibfnamefont {Z.}~\bibnamefont {Yuan}},\ }\bibfield  {title} {\bibinfo
  {title} {High-{Rate} {Measurement}-{Device}-{Independent} {Quantum}
  {Communication} without {Optical} {Reference} {Light}},\ }\href
  {https://doi.org/10.1103/PhysRevX.15.021066} {\bibfield  {journal} {\bibinfo
  {journal} {Physical Review X}\ }\textbf {\bibinfo {volume} {15}},\ \bibinfo
  {pages} {021066} (\bibinfo {year} {2025})}\BibitemShut {NoStop}%
\bibitem [{\citenamefont {Lu}\ \emph {et~al.}(2025{\natexlab{b}})\citenamefont
  {Lu}, \citenamefont {Zhou}, \citenamefont {Guo}, \citenamefont {Li},
  \citenamefont {Jiang}, \citenamefont {Wang}, \citenamefont {Zhou},
  \citenamefont {Li}, \citenamefont {Zhou}, \citenamefont {Li}, \citenamefont
  {Zhou},\ and\ \citenamefont {Bao}}]{mp_lu_experimental_2025}%
  \BibitemOpen
  \bibfield  {author} {\bibinfo {author} {\bibfnamefont {Y.-F.}\ \bibnamefont
  {Lu}}, \bibinfo {author} {\bibfnamefont {Y.-Y.}\ \bibnamefont {Zhou}},
  \bibinfo {author} {\bibfnamefont {Y.-Y.}\ \bibnamefont {Guo}}, \bibinfo
  {author} {\bibfnamefont {X.-H.}\ \bibnamefont {Li}}, \bibinfo {author}
  {\bibfnamefont {X.-L.}\ \bibnamefont {Jiang}}, \bibinfo {author}
  {\bibfnamefont {Y.}~\bibnamefont {Wang}}, \bibinfo {author} {\bibfnamefont
  {Y.}~\bibnamefont {Zhou}}, \bibinfo {author} {\bibfnamefont {J.-J.}\
  \bibnamefont {Li}}, \bibinfo {author} {\bibfnamefont {C.}~\bibnamefont
  {Zhou}}, \bibinfo {author} {\bibfnamefont {H.-W.}\ \bibnamefont {Li}},
  \bibinfo {author} {\bibfnamefont {L.-J.}\ \bibnamefont {Zhou}},\ and\
  \bibinfo {author} {\bibfnamefont {W.-S.}\ \bibnamefont {Bao}},\ }\bibfield
  {title} {\bibinfo {title} {Experimental {Frequency}-{Comb}-{Based}
  {Mode}-{Pairing} {Quantum} {Key} {Distribution} {Beyond} the {Rate}-{Loss}
  {Limit}}\ }\href {https://doi.org/10.48550/arXiv.2505.09223}
  {10.48550/arXiv.2505.09223} (\bibinfo {year} {2025}{\natexlab{b}}),\ \bibinfo
  {note} {arXiv:2505.09223 [quant-ph]}\BibitemShut {NoStop}%
\bibitem [{\citenamefont {Li}\ \emph {et~al.}(2023{\natexlab{b}})\citenamefont
  {Li}, \citenamefont {Zhang}, \citenamefont {Lu}, \citenamefont {Li},
  \citenamefont {Jiang}, \citenamefont {Liu}, \citenamefont {Huang},
  \citenamefont {Li}, \citenamefont {Wang}, \citenamefont {Wang}, \citenamefont
  {Zhang}, \citenamefont {You}, \citenamefont {Xu},\ and\ \citenamefont
  {Pan}}]{li_twin-field_2023}%
  \BibitemOpen
  \bibfield  {author} {\bibinfo {author} {\bibfnamefont {W.}~\bibnamefont
  {Li}}, \bibinfo {author} {\bibfnamefont {L.}~\bibnamefont {Zhang}}, \bibinfo
  {author} {\bibfnamefont {Y.}~\bibnamefont {Lu}}, \bibinfo {author}
  {\bibfnamefont {Z.-P.}\ \bibnamefont {Li}}, \bibinfo {author} {\bibfnamefont
  {C.}~\bibnamefont {Jiang}}, \bibinfo {author} {\bibfnamefont
  {Y.}~\bibnamefont {Liu}}, \bibinfo {author} {\bibfnamefont {J.}~\bibnamefont
  {Huang}}, \bibinfo {author} {\bibfnamefont {H.}~\bibnamefont {Li}}, \bibinfo
  {author} {\bibfnamefont {Z.}~\bibnamefont {Wang}}, \bibinfo {author}
  {\bibfnamefont {X.-B.}\ \bibnamefont {Wang}}, \bibinfo {author}
  {\bibfnamefont {Q.}~\bibnamefont {Zhang}}, \bibinfo {author} {\bibfnamefont
  {L.}~\bibnamefont {You}}, \bibinfo {author} {\bibfnamefont {F.}~\bibnamefont
  {Xu}},\ and\ \bibinfo {author} {\bibfnamefont {J.-W.}\ \bibnamefont {Pan}},\
  }\bibfield  {title} {\bibinfo {title} {Twin-{Field} {Quantum} {Key}
  {Distribution} without {Phase} {Locking}},\ }\href
  {https://doi.org/10.1103/PhysRevLett.130.250802} {\bibfield  {journal}
  {\bibinfo  {journal} {Phys. Rev. Lett.}\ }\textbf {\bibinfo {volume} {130}},\
  \bibinfo {pages} {250802} (\bibinfo {year} {2023}{\natexlab{b}})}\BibitemShut
  {NoStop}%
\bibitem [{\citenamefont {Cao}\ \emph {et~al.}(2015)\citenamefont {Cao},
  \citenamefont {Zhang}, \citenamefont {Lo},\ and\ \citenamefont
  {Ma}}]{Cao_2015}%
  \BibitemOpen
  \bibfield  {author} {\bibinfo {author} {\bibfnamefont {Z.}~\bibnamefont
  {Cao}}, \bibinfo {author} {\bibfnamefont {Z.}~\bibnamefont {Zhang}}, \bibinfo
  {author} {\bibfnamefont {H.-K.}\ \bibnamefont {Lo}},\ and\ \bibinfo {author}
  {\bibfnamefont {X.}~\bibnamefont {Ma}},\ }\bibfield  {title} {\bibinfo
  {title} {Discrete-phase-randomized coherent state source and its application
  in quantum key distribution},\ }\href
  {https://doi.org/10.1088/1367-2630/17/5/053014} {\bibfield  {journal}
  {\bibinfo  {journal} {New Journal of Physics}\ }\textbf {\bibinfo {volume}
  {17}},\ \bibinfo {pages} {053014} (\bibinfo {year} {2015})}\BibitemShut
  {NoStop}%
\bibitem [{\citenamefont {Jiang}\ \emph {et~al.}(2020)\citenamefont {Jiang},
  \citenamefont {Yu}, \citenamefont {Hu},\ and\ \citenamefont
  {Wang}}]{PhysRevResearch.2.043304}%
  \BibitemOpen
  \bibfield  {author} {\bibinfo {author} {\bibfnamefont {C.}~\bibnamefont
  {Jiang}}, \bibinfo {author} {\bibfnamefont {Z.-W.}\ \bibnamefont {Yu}},
  \bibinfo {author} {\bibfnamefont {X.-L.}\ \bibnamefont {Hu}},\ and\ \bibinfo
  {author} {\bibfnamefont {X.-B.}\ \bibnamefont {Wang}},\ }\bibfield  {title}
  {\bibinfo {title} {Sending-or-not-sending twin-field quantum key distribution
  with discrete-phase-randomized weak coherent states},\ }\href
  {https://doi.org/10.1103/PhysRevResearch.2.043304} {\bibfield  {journal}
  {\bibinfo  {journal} {Phys. Rev. Res.}\ }\textbf {\bibinfo {volume} {2}},\
  \bibinfo {pages} {043304} (\bibinfo {year} {2020})}\BibitemShut {NoStop}%
\bibitem [{\citenamefont {Jin}\ \emph {et~al.}(2021)\citenamefont {Jin},
  \citenamefont {Zeng}, \citenamefont {Penty},\ and\ \citenamefont
  {Ma}}]{PhysRevApplied.16.034017}%
  \BibitemOpen
  \bibfield  {author} {\bibinfo {author} {\bibfnamefont {A.}~\bibnamefont
  {Jin}}, \bibinfo {author} {\bibfnamefont {P.}~\bibnamefont {Zeng}}, \bibinfo
  {author} {\bibfnamefont {R.~V.}\ \bibnamefont {Penty}},\ and\ \bibinfo
  {author} {\bibfnamefont {X.}~\bibnamefont {Ma}},\ }\bibfield  {title}
  {\bibinfo {title} {Reference-frame-independent design of phase-matching
  quantum key distribution},\ }\href
  {https://doi.org/10.1103/PhysRevApplied.16.034017} {\bibfield  {journal}
  {\bibinfo  {journal} {Phys. Rev. Appl.}\ }\textbf {\bibinfo {volume} {16}},\
  \bibinfo {pages} {034017} (\bibinfo {year} {2021})}\BibitemShut {NoStop}%
\bibitem [{Note1()}]{Note1}%
  \BibitemOpen
  \bibinfo {note} {\label {footnote}See Supplemental Material for the detailed
  analysis of the frequency difference estimation method (Sec. S3)}\BibitemShut
  {NoStop}%
\bibitem [{\citenamefont {Kippenberg}\ \emph {et~al.}(2018)\citenamefont
  {Kippenberg}, \citenamefont {Gaeta}, \citenamefont {Lipson},\ and\
  \citenamefont {Gorodetsky}}]{kippenberg_dissipative_2018}%
  \BibitemOpen
  \bibfield  {author} {\bibinfo {author} {\bibfnamefont {T.~J.}\ \bibnamefont
  {Kippenberg}}, \bibinfo {author} {\bibfnamefont {A.~L.}\ \bibnamefont
  {Gaeta}}, \bibinfo {author} {\bibfnamefont {M.}~\bibnamefont {Lipson}},\ and\
  \bibinfo {author} {\bibfnamefont {M.~L.}\ \bibnamefont {Gorodetsky}},\
  }\bibfield  {title} {\bibinfo {title} {Dissipative {Kerr} solitons in optical
  microresonators},\ }\href {https://doi.org/10.1126/science.aan8083}
  {\bibfield  {journal} {\bibinfo  {journal} {Science}\ }\textbf {\bibinfo
  {volume} {361}},\ \bibinfo {pages} {eaan8083} (\bibinfo {year}
  {2018})}\BibitemShut {NoStop}%
\bibitem [{\citenamefont {Kues}\ \emph {et~al.}(2019)\citenamefont {Kues},
  \citenamefont {Reimer}, \citenamefont {Lukens}, \citenamefont {Munro},
  \citenamefont {Weiner}, \citenamefont {Moss},\ and\ \citenamefont
  {Morandotti}}]{kues_quantum_2019}%
  \BibitemOpen
  \bibfield  {author} {\bibinfo {author} {\bibfnamefont {M.}~\bibnamefont
  {Kues}}, \bibinfo {author} {\bibfnamefont {C.}~\bibnamefont {Reimer}},
  \bibinfo {author} {\bibfnamefont {J.~M.}\ \bibnamefont {Lukens}}, \bibinfo
  {author} {\bibfnamefont {W.~J.}\ \bibnamefont {Munro}}, \bibinfo {author}
  {\bibfnamefont {A.~M.}\ \bibnamefont {Weiner}}, \bibinfo {author}
  {\bibfnamefont {D.~J.}\ \bibnamefont {Moss}},\ and\ \bibinfo {author}
  {\bibfnamefont {R.}~\bibnamefont {Morandotti}},\ }\bibfield  {title}
  {\bibinfo {title} {Quantum optical microcombs},\ }\href
  {https://doi.org/10.1038/s41566-019-0363-0} {\bibfield  {journal} {\bibinfo
  {journal} {Nature Photonics}\ }\textbf {\bibinfo {volume} {13}},\ \bibinfo
  {pages} {170} (\bibinfo {year} {2019})}\BibitemShut {NoStop}%
\bibitem [{\citenamefont {Shen}\ \emph {et~al.}(2020)\citenamefont {Shen},
  \citenamefont {Chang}, \citenamefont {Liu}, \citenamefont {Wang},
  \citenamefont {Yang}, \citenamefont {Xiang}, \citenamefont {Wang},
  \citenamefont {He}, \citenamefont {Liu}, \citenamefont {Xie}, \citenamefont
  {Guo}, \citenamefont {Kinghorn}, \citenamefont {Wu}, \citenamefont {Ji},
  \citenamefont {Kippenberg}, \citenamefont {Vahala},\ and\ \citenamefont
  {Bowers}}]{shen_integrated_2020}%
  \BibitemOpen
  \bibfield  {author} {\bibinfo {author} {\bibfnamefont {B.}~\bibnamefont
  {Shen}}, \bibinfo {author} {\bibfnamefont {L.}~\bibnamefont {Chang}},
  \bibinfo {author} {\bibfnamefont {J.}~\bibnamefont {Liu}}, \bibinfo {author}
  {\bibfnamefont {H.}~\bibnamefont {Wang}}, \bibinfo {author} {\bibfnamefont
  {Q.-F.}\ \bibnamefont {Yang}}, \bibinfo {author} {\bibfnamefont
  {C.}~\bibnamefont {Xiang}}, \bibinfo {author} {\bibfnamefont {R.~N.}\
  \bibnamefont {Wang}}, \bibinfo {author} {\bibfnamefont {J.}~\bibnamefont
  {He}}, \bibinfo {author} {\bibfnamefont {T.}~\bibnamefont {Liu}}, \bibinfo
  {author} {\bibfnamefont {W.}~\bibnamefont {Xie}}, \bibinfo {author}
  {\bibfnamefont {J.}~\bibnamefont {Guo}}, \bibinfo {author} {\bibfnamefont
  {D.}~\bibnamefont {Kinghorn}}, \bibinfo {author} {\bibfnamefont
  {L.}~\bibnamefont {Wu}}, \bibinfo {author} {\bibfnamefont {Q.-X.}\
  \bibnamefont {Ji}}, \bibinfo {author} {\bibfnamefont {T.~J.}\ \bibnamefont
  {Kippenberg}}, \bibinfo {author} {\bibfnamefont {K.}~\bibnamefont {Vahala}},\
  and\ \bibinfo {author} {\bibfnamefont {J.~E.}\ \bibnamefont {Bowers}},\
  }\bibfield  {title} {\bibinfo {title} {Integrated turnkey soliton
  microcombs},\ }\href {https://doi.org/10.1038/s41586-020-2358-x} {\bibfield
  {journal} {\bibinfo  {journal} {Nature}\ }\textbf {\bibinfo {volume} {582}},\
  \bibinfo {pages} {365} (\bibinfo {year} {2020})}\BibitemShut {NoStop}%
\bibitem [{\citenamefont {Moille}\ \emph {et~al.}(2023)\citenamefont {Moille},
  \citenamefont {Stone}, \citenamefont {Chojnacky}, \citenamefont {Shrestha},
  \citenamefont {Javid}, \citenamefont {Menyuk},\ and\ \citenamefont
  {Srinivasan}}]{moille_kerr-induced_2023}%
  \BibitemOpen
  \bibfield  {author} {\bibinfo {author} {\bibfnamefont {G.}~\bibnamefont
  {Moille}}, \bibinfo {author} {\bibfnamefont {J.}~\bibnamefont {Stone}},
  \bibinfo {author} {\bibfnamefont {M.}~\bibnamefont {Chojnacky}}, \bibinfo
  {author} {\bibfnamefont {R.}~\bibnamefont {Shrestha}}, \bibinfo {author}
  {\bibfnamefont {U.~A.}\ \bibnamefont {Javid}}, \bibinfo {author}
  {\bibfnamefont {C.}~\bibnamefont {Menyuk}},\ and\ \bibinfo {author}
  {\bibfnamefont {K.}~\bibnamefont {Srinivasan}},\ }\bibfield  {title}
  {\bibinfo {title} {Kerr-induced synchronization of a cavity soliton to an
  optical reference},\ }\href {https://doi.org/10.1038/s41586-023-06730-0}
  {\bibfield  {journal} {\bibinfo  {journal} {Nature}\ }\textbf {\bibinfo
  {volume} {624}},\ \bibinfo {pages} {267} (\bibinfo {year}
  {2023})}\BibitemShut {NoStop}%
\bibitem [{\citenamefont {Huang}\ \emph {et~al.}(2025)\citenamefont {Huang},
  \citenamefont {Wang}, \citenamefont {Wang}, \citenamefont {Wang},
  \citenamefont {Zou}, \citenamefont {Tang}, \citenamefont {Little},
  \citenamefont {Zhao}, \citenamefont {Han}, \citenamefont {Yang},
  \citenamefont {Wang}, \citenamefont {Chen},\ and\ \citenamefont
  {Zhang}}]{huang_massively_2025}%
  \BibitemOpen
  \bibfield  {author} {\bibinfo {author} {\bibfnamefont {L.}~\bibnamefont
  {Huang}}, \bibinfo {author} {\bibfnamefont {W.}~\bibnamefont {Wang}},
  \bibinfo {author} {\bibfnamefont {F.}~\bibnamefont {Wang}}, \bibinfo {author}
  {\bibfnamefont {Y.}~\bibnamefont {Wang}}, \bibinfo {author} {\bibfnamefont
  {C.}~\bibnamefont {Zou}}, \bibinfo {author} {\bibfnamefont {L.}~\bibnamefont
  {Tang}}, \bibinfo {author} {\bibfnamefont {B.~E.}\ \bibnamefont {Little}},
  \bibinfo {author} {\bibfnamefont {W.}~\bibnamefont {Zhao}}, \bibinfo {author}
  {\bibfnamefont {Z.}~\bibnamefont {Han}}, \bibinfo {author} {\bibfnamefont
  {J.}~\bibnamefont {Yang}}, \bibinfo {author} {\bibfnamefont {G.}~\bibnamefont
  {Wang}}, \bibinfo {author} {\bibfnamefont {W.}~\bibnamefont {Chen}},\ and\
  \bibinfo {author} {\bibfnamefont {W.}~\bibnamefont {Zhang}},\ }\bibfield
  {title} {\bibinfo {title} {Massively parallel {Hong}-{Ou}-{Mandel}
  interference based on independent soliton microcombs},\ }\href
  {https://doi.org/10.1126/sciadv.adq8982} {\bibfield  {journal} {\bibinfo
  {journal} {Science Advances}\ }\textbf {\bibinfo {volume} {11}},\ \bibinfo
  {pages} {eadq8982} (\bibinfo {year} {2025})}\BibitemShut {NoStop}%
\bibitem [{\citenamefont {Yan}\ \emph {et~al.}(2025{\natexlab{b}})\citenamefont
  {Yan}, \citenamefont {Hu}, \citenamefont {Du}, \citenamefont {Wang},
  \citenamefont {Lu}, \citenamefont {Zhu},\ and\ \citenamefont
  {Ma}}]{yan_ten-channel_2025}%
  \BibitemOpen
  \bibfield  {author} {\bibinfo {author} {\bibfnamefont {W.}~\bibnamefont
  {Yan}}, \bibinfo {author} {\bibfnamefont {Y.}~\bibnamefont {Hu}}, \bibinfo
  {author} {\bibfnamefont {Y.}~\bibnamefont {Du}}, \bibinfo {author}
  {\bibfnamefont {K.}~\bibnamefont {Wang}}, \bibinfo {author} {\bibfnamefont
  {Y.-Q.}\ \bibnamefont {Lu}}, \bibinfo {author} {\bibfnamefont
  {S.}~\bibnamefont {Zhu}},\ and\ \bibinfo {author} {\bibfnamefont {X.-S.}\
  \bibnamefont {Ma}},\ }\bibfield  {title} {\bibinfo {title} {Ten-channel
  {Hong}–{Ou}–{Mandel} interference between independent optical combs},\
  }\href {https://doi.org/10.3788/COL202523.042701} {\bibfield  {journal}
  {\bibinfo  {journal} {Chinese Optics Letters}\ }\textbf {\bibinfo {volume}
  {23}},\ \bibinfo {pages} {042701} (\bibinfo {year}
  {2025}{\natexlab{b}})}\BibitemShut {NoStop}%
\bibitem [{\citenamefont {Wang}\ \emph {et~al.}(2018)\citenamefont {Wang},
  \citenamefont {Zhang}, \citenamefont {Chen}, \citenamefont {Bertrand},
  \citenamefont {Shams-Ansari}, \citenamefont {Chandrasekhar}, \citenamefont
  {Winzer},\ and\ \citenamefont {Lončar}}]{wang_integrated_2018}%
  \BibitemOpen
  \bibfield  {author} {\bibinfo {author} {\bibfnamefont {C.}~\bibnamefont
  {Wang}}, \bibinfo {author} {\bibfnamefont {M.}~\bibnamefont {Zhang}},
  \bibinfo {author} {\bibfnamefont {X.}~\bibnamefont {Chen}}, \bibinfo {author}
  {\bibfnamefont {M.}~\bibnamefont {Bertrand}}, \bibinfo {author}
  {\bibfnamefont {A.}~\bibnamefont {Shams-Ansari}}, \bibinfo {author}
  {\bibfnamefont {S.}~\bibnamefont {Chandrasekhar}}, \bibinfo {author}
  {\bibfnamefont {P.}~\bibnamefont {Winzer}},\ and\ \bibinfo {author}
  {\bibfnamefont {M.}~\bibnamefont {Lončar}},\ }\bibfield  {title} {\bibinfo
  {title} {Integrated lithium niobate electro-optic modulators operating at
  {CMOS}-compatible voltages},\ }\href
  {https://doi.org/10.1038/s41586-018-0551-y} {\bibfield  {journal} {\bibinfo
  {journal} {Nature}\ }\textbf {\bibinfo {volume} {562}},\ \bibinfo {pages}
  {101} (\bibinfo {year} {2018})}\BibitemShut {NoStop}%
\bibitem [{\citenamefont {Grünenfelder}\ \emph {et~al.}(2023)\citenamefont
  {Grünenfelder}, \citenamefont {Boaron}, \citenamefont {Resta}, \citenamefont
  {Perrenoud}, \citenamefont {Rusca}, \citenamefont {Barreiro}, \citenamefont
  {Houlmann}, \citenamefont {Sax}, \citenamefont {Stasi}, \citenamefont
  {El-Khoury}, \citenamefont {Hänggi}, \citenamefont {Bosshard}, \citenamefont
  {Bussières},\ and\ \citenamefont {Zbinden}}]{grunenfelder_fast_2023}%
  \BibitemOpen
  \bibfield  {author} {\bibinfo {author} {\bibfnamefont {F.}~\bibnamefont
  {Grünenfelder}}, \bibinfo {author} {\bibfnamefont {A.}~\bibnamefont
  {Boaron}}, \bibinfo {author} {\bibfnamefont {G.~V.}\ \bibnamefont {Resta}},
  \bibinfo {author} {\bibfnamefont {M.}~\bibnamefont {Perrenoud}}, \bibinfo
  {author} {\bibfnamefont {D.}~\bibnamefont {Rusca}}, \bibinfo {author}
  {\bibfnamefont {C.}~\bibnamefont {Barreiro}}, \bibinfo {author}
  {\bibfnamefont {R.}~\bibnamefont {Houlmann}}, \bibinfo {author}
  {\bibfnamefont {R.}~\bibnamefont {Sax}}, \bibinfo {author} {\bibfnamefont
  {L.}~\bibnamefont {Stasi}}, \bibinfo {author} {\bibfnamefont
  {S.}~\bibnamefont {El-Khoury}}, \bibinfo {author} {\bibfnamefont
  {E.}~\bibnamefont {Hänggi}}, \bibinfo {author} {\bibfnamefont
  {N.}~\bibnamefont {Bosshard}}, \bibinfo {author} {\bibfnamefont
  {F.}~\bibnamefont {Bussières}},\ and\ \bibinfo {author} {\bibfnamefont
  {H.}~\bibnamefont {Zbinden}},\ }\bibfield  {title} {\bibinfo {title} {Fast
  single-photon detectors and real-time key distillation enable high
  secret-key-rate quantum key distribution systems},\ }\href
  {https://doi.org/10.1038/s41566-023-01168-2} {\bibfield  {journal} {\bibinfo
  {journal} {Nature Photonics}\ }\textbf {\bibinfo {volume} {17}},\ \bibinfo
  {pages} {422} (\bibinfo {year} {2023})}\BibitemShut {NoStop}%
\bibitem [{Note2()}]{Note2}%
  \BibitemOpen
  \bibinfo {note} {\protect \url
  {https://github.com/NJU-Malab/AMDI-QCC}}\BibitemShut {NoStop}%
\end{thebibliography}%

\end{document}